\documentclass[12pt]{article}

\usepackage[margin=1in]{geometry}  
\usepackage{epsfig}
\usepackage{graphics,graphicx,color}
\usepackage{amssymb,amsfonts,amsthm,amsmath}
\usepackage{multirow}
\usepackage{mathrsfs}
\usepackage{natbib}
\usepackage{verbatim}
\usepackage{subfigure}
\usepackage{stfloats}
\usepackage{caption}
 
\usepackage{lipsum}       
\usepackage{changepage}
\newenvironment{myenv}{\begin{adjustwidth}{1.25cm}{}}{\end{adjustwidth}}
\newcommand{\bx}{\textbf{x}}
\newcommand{\by}{\textbf{y}}
\newcommand{\bD}{\textbf{D}}
\newcommand{\bone}{\textbf{1}}
\newcommand{\bR}{\textbf{R}}
\newcommand{\br}{\textbf{r}}

\newcommand{\bbeta}{\mbox{\boldmath${\beta}$} }
\newcommand{\bff}{\textbf{f}}

\def\TT{{\mbox{\tiny T}}}

\newcommand{\pkg}[1]{{\normalfont\fontseries{b}\selectfont #1}} 
\let\proglang=\textit
\let\code=\texttt

\begin{document}

\baselineskip=22pt \vskip 20pt
\begin{center}
{\Large \bf Global Fitting of the Response Surface via Estimating Multiple Contours of a Simulator}
\\
\vskip 20pt
{\bf F.\ Yang}$^{\dag}$, {\bf C.\ Devon Lin}$^{\ddag}$\footnote{Address for correspondence: C.\ Devon Lin, Associate Professor, Department of Mathematics and Statistics,
Queen's University, Kingston, ON Canada (E-mail: devon.lin@queensu.ca).},  and {\bf P. Ranjan}$^{\S}$ \\[0.1in]
$^\dag$College of Mathematics, Sichuan University, China \\
$^\ddag$Department of Mathematics and Statistics,
Queen's University, Canada \\
$^\S$OM\&QT Area, 
    Indian Institute of Management Indore, India\\ 
\end{center}

\baselineskip=16pt \vskip 10pt

{\bf Abstract}
Computer simulators are nowadays widely used to understand complex physical systems in many areas such as aerospace, renewable energy, climate modeling, and manufacturing. One fundamental issue in the study of computer simulators is known as experimental design, that is, how to select the input settings where the computer simulator is run and the corresponding response is collected.  Extra care should be taken in the selection  process because computer simulators can be computationally expensive to run. The selection shall acknowledge and achieve the goal of the analysis. This article focuses on the goal of producing more accurate prediction which is important for risk assessment and decision making.  We propose two new methods of design approaches that sequentially select input settings to achieve this goal. The approaches make novel applications of simultaneous and sequential contour estimations. Numerical examples are employed to demonstrate the effectiveness of the proposed approaches. \\


\baselineskip=22pt \vskip 20pt

\section{Introduction}

Computer models or simulators are increasingly becoming popular for gaining insights of the physical processes and phenomena that are too expensive or infeasible to observe. For example, Greenberg (1979) developed a finite volume community ocean model (FVCOM) for simulating the flow of water in the Bay of Fundy; Bower et al. (2006) discussed the formation of galaxies using a simulator called GALFORM; and Bayarri et al. (2009) used a simulator called TITAN2D for modelling the maximum volcanic eruption flow height. Realistic computer simulators of complex processes can also be computationally expensive to run, and thus statistical surrogates trained on a handful of simulator runs are often used for the deeper understanding of the underlying phenomena. Sacks et al. (1989) proposed using a realization  of the Gaussian process (GP) model as a surrogate for such processes.

The popular objectives of such computer experiments include global fitting, variable screening, and estimation of process features like the maximum, a pre-specified contour or a tail quantile region. Assuming the simulator under consideration is expensive to run, the number of simulator runs would be limited and thus one must be careful in choosing the inputs. Over the last two decades several innovative methodologies and algorithms have been developed to address some of the concerns. See Santner, Williams and Notz (2003), Fang, Li and Sudjianto (2005) and Rasmussen and Williams (2006) for details.

We focus on efficient designs for global fitting. In computer experiments literature, a popular technique is to use Latin hypercube designs (McKay et al., 1979) with some space-filling properties like maximin interpoint distance (Johnson et al., 1990, Morris and Mitchell, 1995), minimum pairwise coordinate correlation (Iman and Conover, 1982, Joseph and Hung, 2008), orthogonal array-based structure (Owen, 1992, Tang, 1993), projection property (Joseph, Gul and Ba, 2015), etc.  Such designs aim at filling the input space as evenly as possible, but do not consider the complexity of the response surface. On the other hand, D-optimal designs (Johnson et al., 1990), integrated mean squared prediction error (IMSPE)-optimal designs  (Sacks, Schiller and Welch, 1989) and maximum mean squared prediction error (MMSPE)-optimal designs (Sacks and Schiller, 1988) use the process response information in finding a design for global fitting.

Most of these designs follow one-shot approach, i.e., all design points are obtained at the same time. However, over the past decade, a few sequential designs have also been proposed for global fitting of the response surface that have higher prediction accuracy. For instance, the D-optimal design (Gramacy and Lee, 2009), expected improvement (EI) criterion based design (Lam and Notz, 2008) and minimum potential energy based design (Joseph et al., 2015). In this paper, we propose two new sequential design approaches for global fitting.

The main idea behind our proposed approaches comes from the fact that the estimation of a response surface can be approximated by the estimation of a large number of contours over the range of the responses.  This further motivated us to generalize the contour estimation idea (Ranjan et al., 2008) for our objective. In this paper, we propose two generalizations. First, we recommend splitting the range of simulator outputs into $k$ equi-spaced contours and then develop a new EI criterion for the simultaneous estimation of these pre-specified multiple contours. Second, we propose a new adaptive approach of choosing contour levels for selecting the follow-up trial by maximizing the EI criterion for contour estimation. The performance of the proposed approaches have been compared with several state of the art designs for global fitting.

The remainder of the article is organized as follows. Section~2 presents a quick review of the Gaussian process (GP) model for building a surrogate of the computer model output, popular sequential design approaches for global fitting (Lam and Notz, 2008, Joseph et al., 2015) and the EI criterion for contour estimation (Ranjan et al., 2008). Section~3 presents the new multiple contours estimation-based EI method for constructing designs for global fitting of the response surface. In Section~4, we propose the new adaptive method of estimating the contour levels for choosing follow-up design points in the sequential framework. The performance comparison of the proposed methods and the existing approaches are discussed in Section~5. Finally, Section~6 summarizes the key findings and concluding remarks.


\section{Background Review}

This section reviews the necessary background and the existing relevant work for later development. More specifically, we provide a brief account of reviews on Gaussian process models used throughout, the existing sequential design approaches for global fitting as well as the contour estimation in Ranjan et al. (2008).   Although these topics can easily be accessed in the literature, we include them here so that this article is a standalone document. 


\subsection{Gaussian Process Models }

Gaussian process models are most widely used in computer experiments to emulate outputs from computer codes (e.g., Sacks et al., 1989). Its popularity is due to its simplicity, flexibility and the ability of providing the predictive uncertainty. Here we cover the key concepts of GP models  and refer the reader to Santner, Williams and Notz (2003) and Rasmussen and Williams (2006) for details.  For a training data of size $n$, let the $i$th input and output of a computer code be a $d$-dimensional vector $\bx_i =(x_{i1}, \ldots, x_{id})$ and a scalar  $y_i = y(\bx_i)$, for $i=1,\ldots,n$. Typically, without the loss of generality, the design domain is assumed to be a unit hypercube,  $\chi = (0,1)^d$. A GP model assumes 
\begin{equation}\label{eq:gp}
y(\bx_i) = \bff \bbeta+ Z(\bx_i), i= 1,2,...,n,
\end{equation}
\noindent where $\bff$ is a vector of regression functions, $\bbeta$ is the vector of regression parameters, 
 $Z(\bx)$ is a stationary stochastic process with mean zero, constant variance $\sigma^2$, and the correlation between two outputs $y(\bx_i)$ and $y(\bx_j)$ being denoted by $R(\bx_i,\bx_j)= \hbox{corr}(\bx_i,\bx_j)$.  In this article, we focus on the Gaussian process models with a constant mean, that is, $\bff \bbeta = \mu$. Let $\by = (y_1,\ldots, y_n)^T$ be the vector of responses for the training data and $\bR$ be an $n\times n$ spatial correlation matrix with the $(i,j)$th element $R(\bx_i, \bx_j)$. A GP model in (\ref{eq:gp}) is equivalent to assume that $\by$ follows a multivariate normal distribution with mean vector $\mu \bone_n$ and the covariance matrix $\sigma^2\bR$ with $\bR=(\bR_{ij})$, where $\bone_n$ is an $n$-dimensional column vector of all 1's. Notationally, we denote $\by \sim GP(\mu \bone_n, \sigma^2\bR)$. There are many choices of valid correlation functions. One popular choice is the Gaussian correlation function, 
\begin{eqnarray}
R(\bx_i, \bx_j) & = &   \prod_{k=1}^d \hbox{exp} \{ -\theta_k (x_{ik} - x_{jk})^2\},
\end{eqnarray} 
where $\theta_k$ is the correlation parameter for the $k$th input variable.  The unknown parameters in the model include the mean $\mu$, the variance $\sigma^2$, and $d$ correlation parameters $\theta_1, \ldots, \theta_d$. 
 They can be estimated via the maximum likelihood approach or Bayesian approach such as Markov Chain Monte Carlo (MCMC) (Santner et al., 2003, Fang et al., 2005, Currin et al., 1988, Linkletter et al., 2006).  For the maximum likelihood approach, if the correlation parameters are known, the estimates of $\mu$ and $\sigma^2$ in (\ref{eq:gp}) are
 \begin{equation}\label{eq:mu}
 \hat{\mu}  = (\bone_n^\TT\bR^{-1}\bone_n)^{-1}\bone_n^\TT\bR^{-1}\by
 \end{equation} 
 and
 \begin{equation} \label{eq:sigma2}
 \hat{\sigma}^2 = \frac{ (\by-\bone_n\hat{\mu} )^T \bR^{-1} (\by-\bone_n\hat{\mu} )}{n}. 
 \end{equation} 
The best linear unbiased predictor (BLUP) at an input $\bx_0$ is given by
\begin{equation}\label{eq:ok}
\hat{y}(\bx_0) = E[y(\bx_0)|\by] = \mu + \br^T(\bx_0) \bR^{-1}(\by - \mu\bone_n),
\end{equation}
where  $\br(\bx_0) = (R(\bx_0, \bx_1),\ldots,R(\bx_0, \bx_n))^\TT$. 
Moreover, the predictive variance of $y(\bx)$ is  
\begin{equation}\label{eq:var}
s^2(x_0) = \hbox{Var}(y(\bx_0)|{\by}) = \sigma^2 \left(1-\br^T(\bx_0)\bR^{-1}\br(\bx_0) \right).
\end{equation} 
In practice, the unknown correlation parameters in (\ref{eq:mu}) and (\ref{eq:sigma2}) are replaced with the estimates. Thus, $\mu$, $\sigma^2$, $\bR$ and $\br(\bx_0)$  in (\ref{eq:ok}) and (\ref{eq:var}) are replaced by $\hat{\mu}$, $\hat{\sigma}^2$, $\hat{\bR}$ and $\hat{\br}(\bx_0)$, respectively. There are a number of \proglang{R} packages that can provide the GP model fitting. They include \pkg{mlegp}, \pkg{GPfit}, \pkg{DiceKriging}, \pkg{tgp}, \pkg{RobustGaSP} and \pkg{SAVE} (Dancik, 2018, MacDonald, Ranjan and Chipman, 2015, Roustant, Ginsbourger and Deville, 2018, Gramacy and Taddy, 2016, Gu, Palomo and Berger, 2018, Palomo, Paulo and Garcia-Donato, 2015). These \proglang{R} packages are different in terms of computational efficiency and stability. In general they shall provide similar results. For the reason of stability, we use the \proglang{R} package \pkg{GPfit} (MacDonald et al., 2015) in this article. 
 

\subsection{Existing Sequential Design Approaches for Global Fitting}

The general setup of a sequential design approach starts with an initial design and adds one point or a batch of points at-a-time sequentially. We focus on the sequential approaches of adding one point at-a-time. The next follow-up point shall be chosen based on the information gathered from the existing data and shall be most informative among the candidate points. The process of adding points is repeated until a tolerance based stopping criterion is met or a pre-specified  budget is exhausted. The step-by-step process of the sequential design approach is as follows. 
\begin{myenv}
\begin{itemize}
    \item[Step 1.] Choose an initial design of run size $n_0$.  Let $n=n_0$. 
    \item[Step 2.] Build a statistical surrogate model using the available data $\{ (\bx_i, y_i), i = 1,\ldots,n\}$. 
    \item[Step 3.] Choose the next design point $\bx_{n+1}$ based on a criterion. Run the computer code at the new input 
    $\bx_{n+1}$ and obtain the corresponding response $y_{n+1}$.  
   \item[Step 4.] Let $n= n+1$ and repeat Steps 2 and 3 until it reaches the run size budget or satisfies the stopping criterion. 
 \end{itemize}
\end{myenv}

A few remarks are in order. First, the initial design typically comes with some space-filling property like maximin interpoint distance, minimum pairwise coordinate correlation, etc. If the initial run size $n_0$ is too small, the resulting surrogate model could be wildly inaccurate and mislead the follow-up design choice.  On the other hand, if the $n_0$ is relatively large, it may not fully take the advantage of sequential design criterion in Step~3. The notion of expected improvement (EI) criterion has become extremely popular for choosing follow-up design points after Jones et al. (1998) developed an EI criterion for finding the global minimum of a computer simulator response. One recommendation, given by Ranjan et al. (2008), for the value of $n_0$ is $25-35\%$  of the ultimate run size budget. Such a recommendation is based on their sequential design approach for contour estimation. Second, the run size budget certainly depends on the computer code of interest. 
Loeppky et al. (2009) provided a rule of thumb for selecting a sample size, that is, 10 times the number of input variables.  In our illustrative examples, the total runsize is at least $10d$. Third, in principle, any modelling methods such as GP, treed GP (TGP), or Bayesian additive regression trees (BART) (Gramacy and Lee, 2008, Chipman et al., 2012) can be used as a surrogate in Step~2. We focus on GP modelling in the examples. 


\subsubsection{Expected improvement criterion by Lam and Notz (2008)}

Lam and Notz (2008) introduced a sequential design approach based on an {\em expected improvement for global fit} (EIGF) criterion which chooses the next input point that maximizes the expected improvement 
\begin{equation} \label{eq:eigf}
\hbox{E}(I(\bx))  = (\hat{y}(\bx) - y(\bx_{j^*}))^2 + \hbox{Var}(\hat{y}(\bx)) =   (\hat{y}(\bx) - y(\bx_{j^*}))^2  + s^2(\bx)
\end{equation}
\noindent where the improvement function $I(\bx)$ is defined as 
$$I(\bx) =  (\hat{y}(\bx) - y(\bx_{j^*}))^2 $$
with $y(\bx_{j^*})$ being the observed output at the sampled point, $\bx_{j^*}$, that is closest in distance to the candidate point $\bx$. They use the Euclidean distance to determine this nearest sampled design point. The expectation in (\ref{eq:eigf}) is taken with respect to the predictive distribution of $y(\bx)$ under the GP model, i.e., $y(\bx) \sim N(\hat{y}(\bx), s^2(\bx))$.  The EIGF criterion in (\ref{eq:eigf}) balances the local search and global search of the next potential design input that guides the search for the ``informative'' regions with significant variation in the response values. 
 

\subsubsection{Sequential minimum energy designs by Joseph et al. (2015)}

Motivated by the fact in physics that the charged particles in a box repel and try to remain away from each other as much as possible, 
Joseph et al. (2015) viewed a space-filling design in the experimental region as the positions occupied by the charged particles in a box. The charge of each particle represents the experimental response. A minimum energy design is obtained by minimizing the potential energy.  Let $q(\bx)$ be the charge of the particle at the design input $\bx$ and $d(\bx_i, \bx_j)$ denote the Euclidean distance between the $i$th and the $j$th input. Joseph et al. (2015)  defined the potential energy of a design $\bD_n = \{\bx_1,\ldots, \bx_n\}$ as
\begin{equation}\label{eq:med}
 \hbox{GE}_p =  \left\{ \sum_{i=1}^{n-1} \sum_{j=i+1}^n \left(\frac{q(\bx_i)q(\bx_j)}{d(\bx_i,\bx_j)}\right)^p  \right\}^{1/p},
\end{equation} 
\noindent where $p$ is in the range of $[1, \infty)$. They further proposed a sequential minimum energy design approach which works as follows.  Let $\hat{q}(\bx) = \{ \hat{y}(\bx) \}^{-1/(2d)}$, where $d$ is the dimensionality of the input $\bx$. Then the proposed  one point at-a-time greedy algorithm finds the next follow-up design point given by
\begin{equation}\label{eq:smed}
\bx_{n+1} = \underset{\bx_0 \in \chi}{\arg \min} \sum_{i=1}^n \left( \frac{ \hat{q}(\bx_i)\hat{q}(\bx_0)}{ d(\bx_i, \bx_0)}\right)^p.
\end{equation}
The design generated by this algorithm is called {\em sequential minimum energy design} (SMED).


\subsection{Contour Estimation  via  EI Criterion}

The contour at level ``$a$'' of a simulator response surface consists of all the inputs $\bx$ that yield the same response $a$, that is,
\begin{equation} \label{eq:contour}
S(a)  = \{ \bx\in \chi : y(\bx) = a \}. 
\end{equation}
Ranjan et al. (2008) developed an expected improvement criterion under the sequential design  methodology for estimating a contour from an expensive to evaluate computer simulator with scalar responses. The proposed improvement function is, 
\begin{equation} \label{eq:imp}
I(\bx) = \epsilon^2(\bx) - \min \big \{ (y(\bx) - a)^2, \epsilon^2(\bx) \big \},
\end{equation} 
where $y(\bx)$ has a normal predictive distribution, i.e., $y(\bx) \sim N(\hat{y}(\bx), s^2(\bx))$, and $\epsilon(\bx) = \alpha s(\bx)$ for a positive constant $\alpha$. A suggested value for $\alpha$ is 1.96 for the reason that this value defines a region of interest around $S(a)$ to be 95\% confidence interval under the normality assumption of the responses.  Letting $v_1(\bx) = a - \epsilon(\bx)$ and $v_2(\bx) = a + \epsilon(\bx)$,  the closed form of the expectation of the improvement function $I(\bx)$ with respect to the predictive distribution of $y(\bx)$ is given by,
\begin{eqnarray}\label{eq:ei_original}
\hbox{E}[I(\bx)] & = & \int_{v_1(\bx)}^{v_2(\bx)} [ \epsilon^2(\bx) - (t - a)^2 ] \phi\left(\frac{t-\hat{y}(\bx)}
{s(\bx)}\right) dt \nonumber\\
& =  &
[\epsilon(\bx)^2 - (\hat{y}(\bx) - a)^2 - s^2(\bx) ] ( \Phi(u_2) - \Phi(u_1)) + s^2(\bx) (u_2\phi(u_2) - u_1\phi(u_1)) \nonumber \\
& &  \ \ \ \ \ \ \ \ + 2(\hat{y}(\bx) - a) s(\bx) (\phi(u_2) - \phi(u_1)),
\end{eqnarray}
where $u_1 = [v_1(\bx) - \hat{y}(\bx)]/s(\bx)$, $u_2 = [v_2(\bx) - \hat{y}(\bx)]/s(\bx)$, and $\phi(\cdot)$ and $\Phi(\cdot)$ are the probability density function and the cumulative distribution function of a standard normal random variable, respectively.  See Ranjan et al. (2008) and the associated Errata for the derivation of (\ref{eq:ei_original}). The first term in  (\ref{eq:ei_original})  suggests an input with a large $s(\bx)$ in the neighbourhood of the predicted contour, while  the last term assigns the weights to points that are far away from the predicted contour with large uncertainties. The second term is often dominated by the other two terms in (\ref{eq:ei_original}).  Maximizing the EI criterion in  (\ref{eq:ei_original})  results in the inputs with high uncertainty near the predicted contour as well as those far away, achieving both aims of local search and global exploration.


\section{Global Fitting by Estimating Multiple Contours}

This section proposes a new method for constructing a sequential design for achieving higher prediction accuracy of the overall global fit. The basic sequential framework would remain the same as in Section~2.2, that is, start with a good initial design (e.g., maximin Latin hypercube) of size $n_0 \ll n$ and then sequentially add the remaining $n-n_0$ points using some method that feeds on the objective of global fitting. Instead of the conventional approach of trying to evenly fill the input space, the proposed idea is to slice the response surface into multiple contours and then use the sequential design approach to simultaneously estimate those contours. Next, we generalize the EI criterion for contour estimation (Ranjan et al., 2008) for simultaneous estimation of multiple contours.

For a given integer $k>0$ and the set of scalar values $a_1, \ldots, a_k \in [y_{min}, y_{max}]$, suppose that we are interested in estimating $k$ contours $S(a_1),\ldots,S(a_k)$, where $[y_{min}, y_{max}]$ represents the range of the true simulator response, and $S(\cdot)$ is defined in (\ref{eq:contour}). Without loss of generality, assume
$a_1< a_2 < \cdots <a_k$. For choosing the follow-up trial, we propose the improvement function  at input $\bx$ as
\begin{equation}\label{eq:imp}
I(\bx) = \epsilon^2(\bx) - \min \{ (y(\bx) - a_1)^2,
\ldots, (y(\bx) - a_k)^2, \epsilon^2(\bx) \},
\end{equation}
where $y(\bx) \sim N(\hat{y}(\bx), s^2(\bx))$ and $\epsilon(\bx) = \alpha s(\bx)$ for some positive constant $\alpha$. This improvement function will be non-zero only if $(y(\bx)-a_j)^2 < \epsilon^2(\bx)$ for some $j$. Therefore,  the improvement function can be re-written as:  
\begin{eqnarray*}
I(\bx) &=&  \max\left\{0, \epsilon^2(\bx) - (y(\bx)-a_j)^2,\ \  j=1,2,...,k \right\}. 
\end{eqnarray*}

Since $a_1< a_2 < \cdots <a_k$, the improvement function can be further simplified as 
 \begin{equation*} 
I(\bx) = \left\{
  \begin{array}{ll}
    \epsilon^2(\bx) - (y(\bx) - a_1)^2, & \ \ \
a_1 - \epsilon(\bx) \leq y(\bx) \leq \min\{a_1+\epsilon(\bx), (a_1+a_2)/2\}\\
\ldots & \\
 &  \ \ \ \max\{a_j - \epsilon(\bx), (a_{j-1}+a_j)/2 \} \leq y(\bx), \\
 \epsilon^2(\bx) - (y(\bx) - a_j)^2, & \ \ \ y(\bx) \leq \min\{a_j + \epsilon(\bx), (a_j+a_{j+1})/2\}, \\
&  \ \ \  2 \leq j \leq k-1;
 \\
\ldots & \\
 \epsilon^2(\bx) - (y(\bx) - a_k)^2, & \ \ \
\max\{a_k - \epsilon(\bx), (a_{k-1}+a_k)/2 \} \leq y(\bx) \leq
a_k + \epsilon(\bx),  \\
    0, & \ \ \ \hbox{otherwise}.
  \end{array}
\right.
 \end{equation*}
The term $\epsilon(\bx)$ defines an uncertainty band around each contour that is a function of the predictive standard deviation $s(\bx)$. For the design points already chosen, the radius of the band is exactly zero. In addition, the criterion will tend to be large for the samples from the sets $( \{ \bx:y(\bx) = a_1\},  \{ \bx:y(\bx) = a_2\}, \ldots,
 \{ \bx:y(\bx) = a_k\})$, where $s(\bx)$ is large. 
 
Similar to other sequential design approaches, we suggest choosing follow-up design points by maximizing the corresponding expected improvement, where the expectation is taken with respect to the predictive distribution, $y(\bx) \sim N(\hat{y}(\bx), s^2(\bx))$. For $j=1,\ldots,k$, let $v_{j1}(\bx)$'s and $v_{j2}(\bx)$'s be defined as follows, 
\begin{equation}\label{eq:v1}
v_{j1} (\bx)= \left\{
  \begin{array}{ll}
    a_1 - \epsilon(\bx), & j = 1; \\
   \max\{a_j - \epsilon(\bx), (a_{j-1}+a_j)/2 \} , & 2 \leq j \leq k,\\
  \end{array}
\right.
\end{equation}
and
\begin{equation}\label{eq:v2}
v_{j2}(\bx)= \left\{
  \begin{array}{ll}
    \min\{a_j+\epsilon(\bx), (a_{j}+a_{j+1})/2\}, &  1 \leq j \leq k-1; \\
    a_k + \epsilon(\bx), & j=k.\\
  \end{array}
\right.
\end{equation}
Then, the expectation of the improvement function in (\ref{eq:imp}) is simply the sum of the individual contour estimation EI criterion of Ranjan et al. (2008) over $k$ cases, i.e., 
\begin{eqnarray} \label{eq:ei_mc}
E[I(\bx)] &= &  \sum_{j=1}^k \int_{v_{j1}(\bx)}^{v_{j2}(\bx)}
[ \epsilon^2(\bx) - (t - a_j)^2 ] \phi\left(\frac{t-\hat{y}(\bx)} {s(\bx)}\right) dt \nonumber\\
& = & 
\sum_{j=1}^k \Big \{ [\epsilon(\bx)^2 - (\hat{y}(\bx) - a_j)^2 - s^2(\bx) ] ( \Phi(u_{j2}) - \Phi(u_{j1})) \\
& &  \ \ \ \  \  + s^2(\bx) (u_{j2}\phi(u_{j2}) - u_{j1}\phi(u_{j1})) + 2(\hat{y}(\bx) - a _j) s(\bx) (\phi(u_{j2}) - \phi(u_{j1})) \Big \}, \nonumber
\end{eqnarray}
where $u_{j1} = (v_{j1}(\bx) - \hat{y}(\bx))/s(\bx)$ and $u_{j2} = (v_{j2}(\bx) - \hat{y}(\bx))/s(\bx)$, $\phi(\cdot)$ and $\Phi(\cdot)$ are the probability density function and the cumulative distribution function of a standard normal random variable, respectively.  The formulation in  (\ref{eq:ei_mc}) reduces to (\ref{eq:ei_original}) when the number of contour levels is $k=1$.

Note that the maximization of $E[I(\bx)]$ over $I(\bx)$ has two advantages. First, the true value of $y(\bx)$ (and hence $I(\bx)$) is unknown for any unsampled design point. Second, some regions of the design space may not have been sufficiently explored yet and the predictive variance of $\hat{y}(\bx)$ is relatively high. For such an unsampled design point,  even though the predicted response is not within the $\epsilon(\bx)$-band of one of those $k$ contours, the true contours may lie in the unexplored region. As a result, the EI approach facilitates a balance between the local exploitation versus global exploration.

 A naive way to choose $k$ - the number of contours, and contour levels, $a_1, a_2, \ldots, a_k$, for global fitting is to use equi-spaced $k$ contours in the simulator output range $[y_{min}, y_{max}]$, and finding their optimal values appear to be a challenging task. We now present two illustrations of the proposed multiple contour estimation EI criterion (referred to as MC criterion) for global fitting with different values of $k$.

\begin{figure}[h!]
\centering
\subfigure[$k= 5$]{\label{fig:a} \includegraphics[width= 0.45\textwidth]{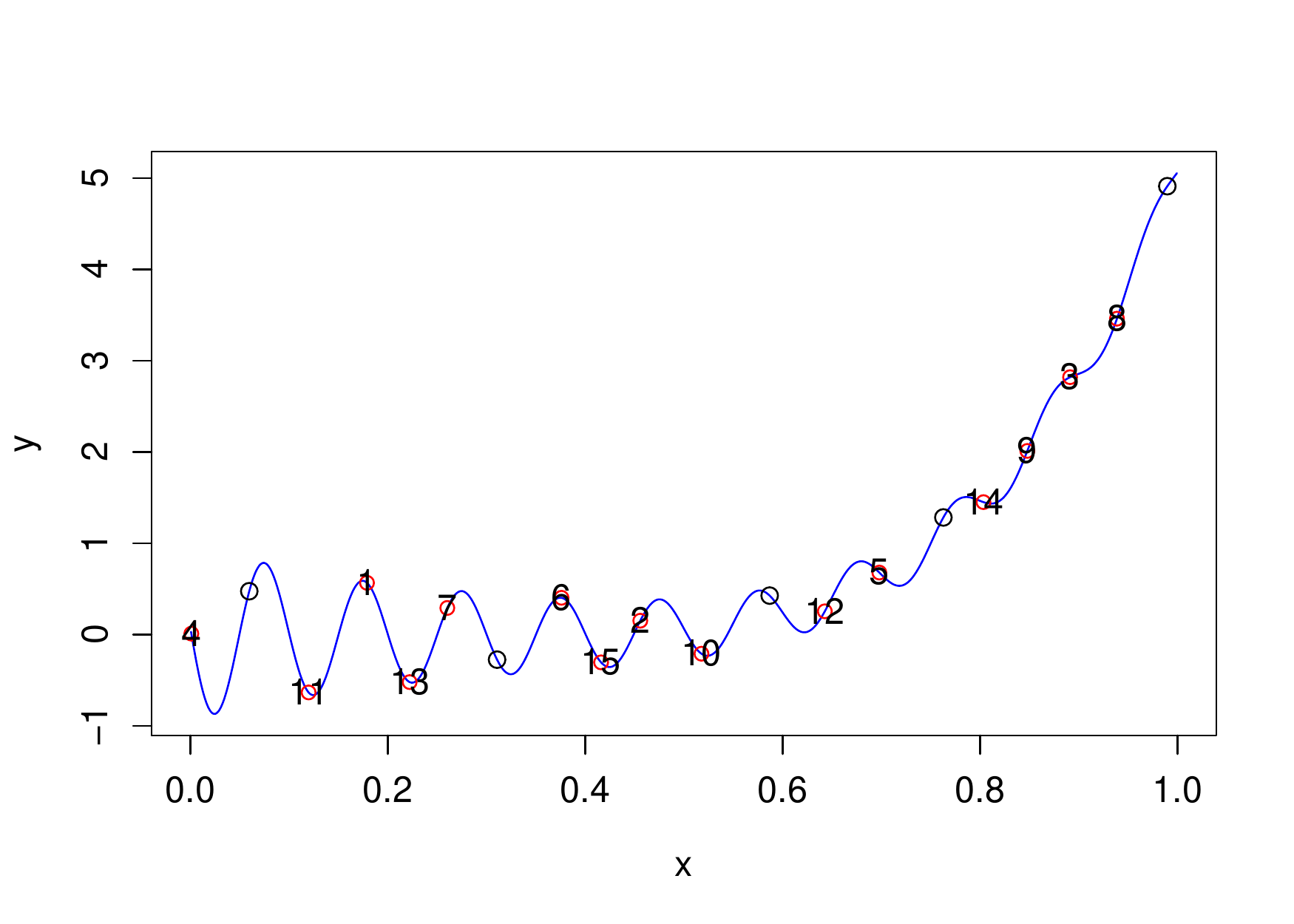} }
\subfigure[$k=20$]{\label{fig:b}  \includegraphics[width=0.45\textwidth]{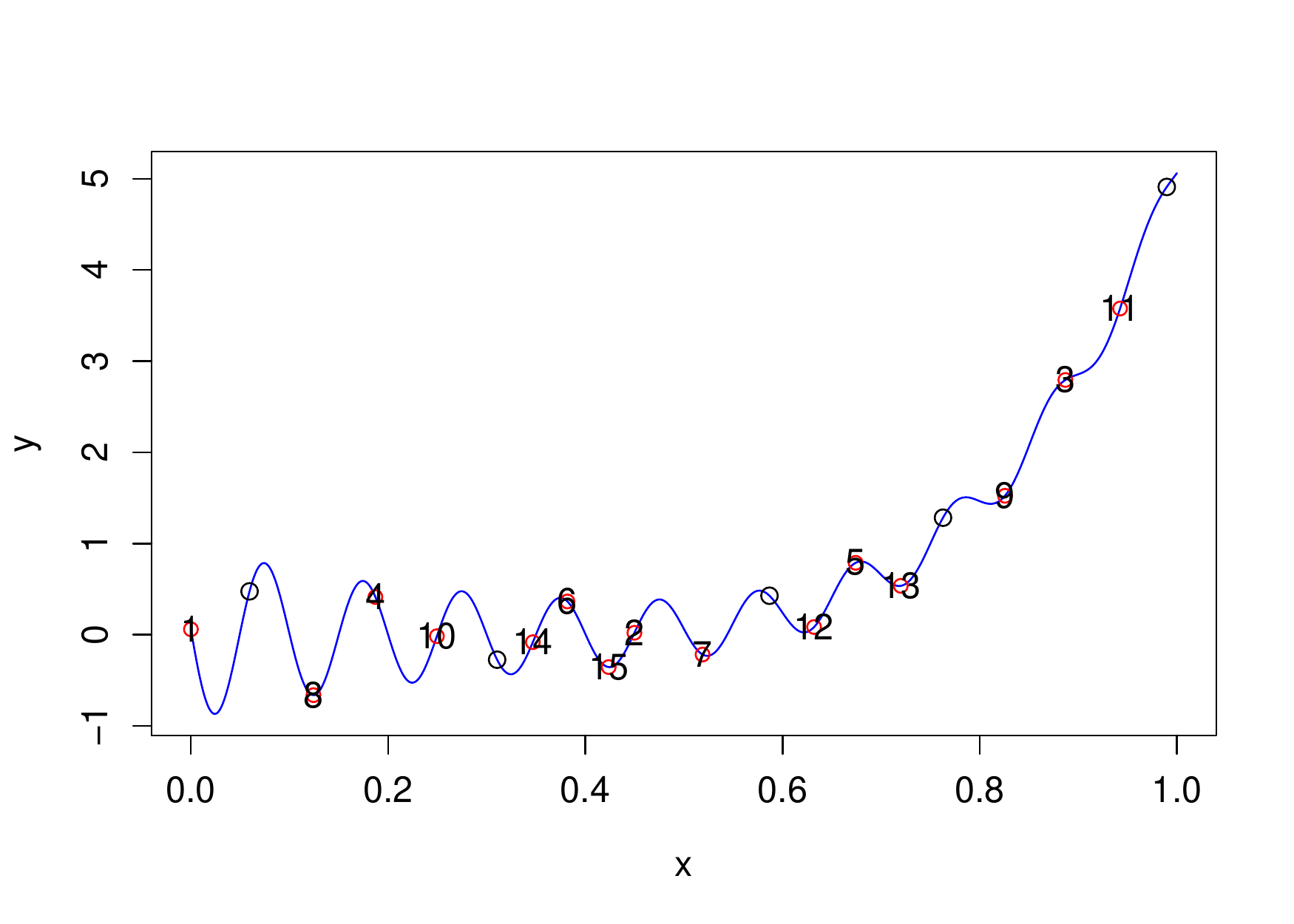}}
\subfigure[$k=50$]{\label{fig:c}  \includegraphics[width=0.45\textwidth]{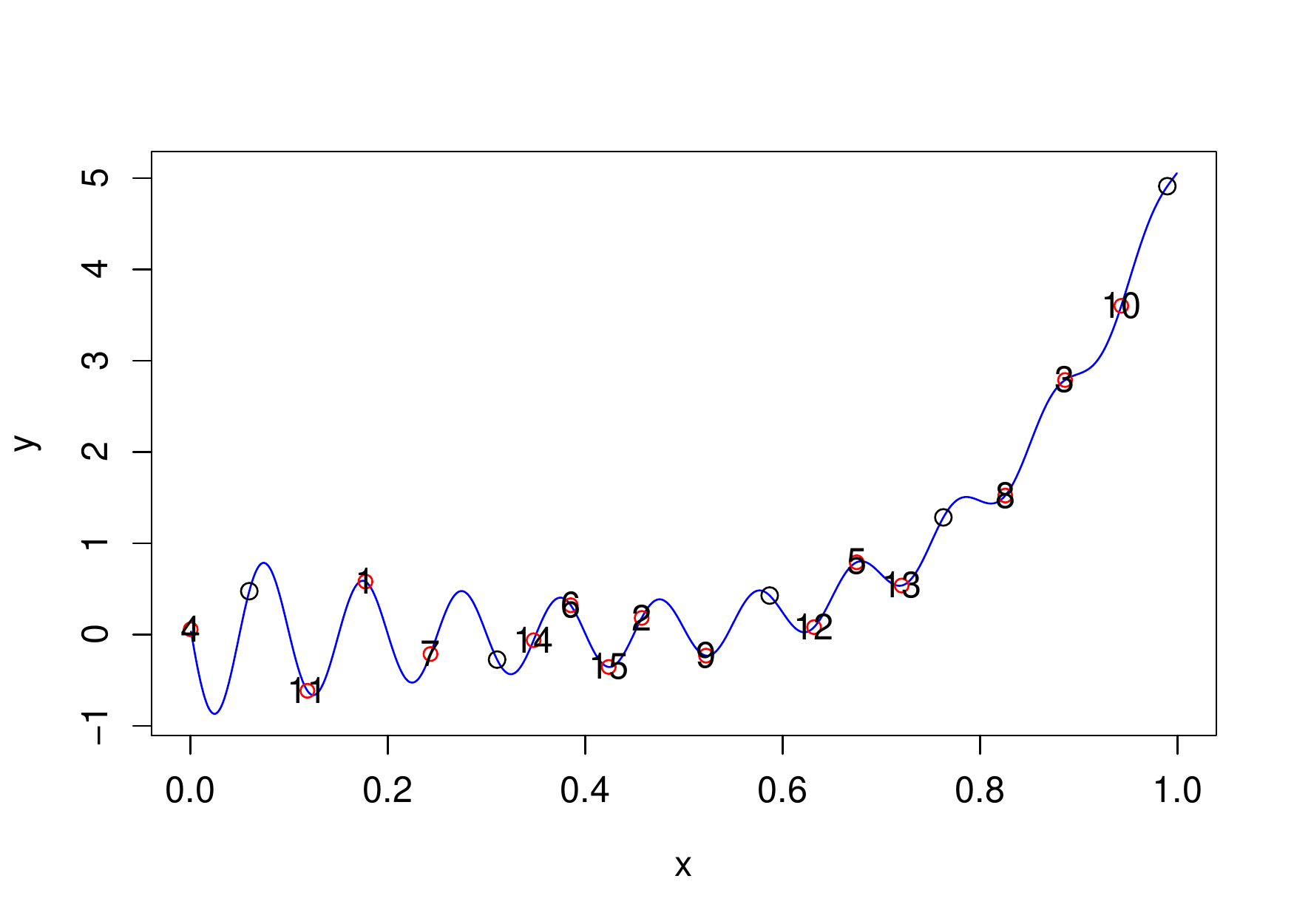}}
\caption{Illustration of MC criterion with $k$ contour levels. The blue curves represent the true relationship between $x$ and $y$ of  the computer model in (\ref{eq:gl}); black empty circles are the  five initial design points; the red numerical labels are locations of follow-up design points.}\label{fig:gl}
\end{figure}

\textbf{Example 1}. Consider the computer model (Gramacy and Lee, 2012) that relates the one-dimensional input $x$ and the output $y$ as, 
\begin{equation}\label{eq:gl}
y = \frac{\hbox{sin}(10\pi x)}{2x} + (x-1)^4, \  \  \ 0.5 \leq x \leq 2.5.
\end{equation}
The true relationship between the input $x$ and the output $y$  is displayed in the blue solid curve in Figure \ref{fig:gl}.  Five initial design points are shown by black empty circles. We then sequentially add 15 design points using the MC criterion in (\ref{eq:ei_mc}). The numerical labels represent the order of the newly added design points. Figures \ref{fig:a},
\ref{fig:b}, and \ref{fig:c} illustrate the sequential design scheme with the MC criterion for 5, 20, 50 equally spaced contour levels within the ranges of the fitted surface.

Clearly Figure \ref{fig:gl} reveals that the proposed MC criterion-based sequential design approach can choose the inputs that are around the areas where the function changes the direction and locate most of the points in the areas where the computer model is more complex. The results show that the final design is in general space-filling but with some nearby points, for example, points 10 and 15, in Figure \ref{fig:c}.

\begin{figure}[h!] \centering
\includegraphics[width=0.6\textwidth]{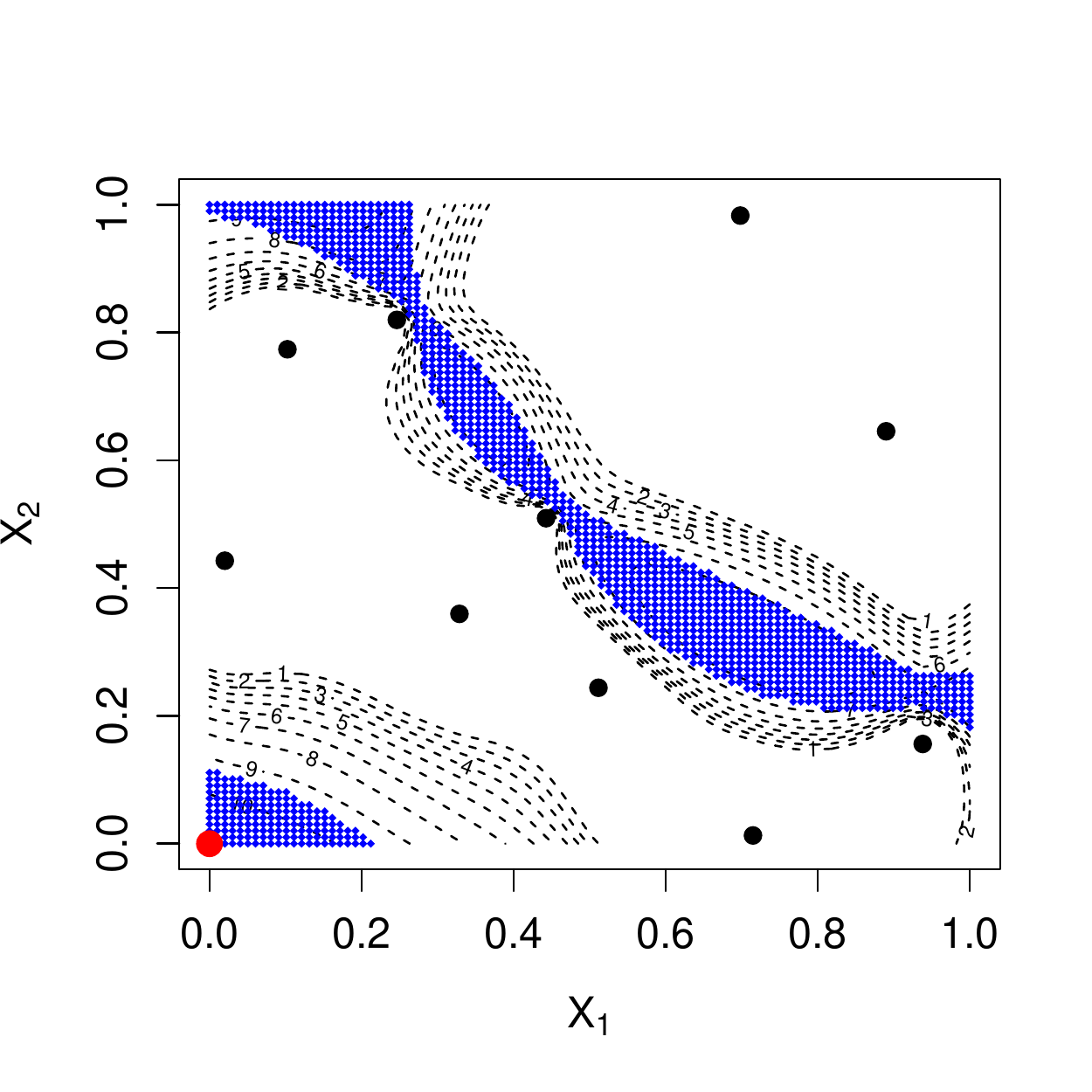}
\caption{Illustration of the follow-up point selection method using the EI criterion for contour estimation from the computer model (\ref{eq:2d}). The black solid circles denote the training points, blue dots represent non-zero improvement value, i.e., $\{\bx: |y(\bx)-a| \le \epsilon(\bx)\}$ for the contour level $a=300$, the contour lines display $\log(E[I(\bx)])$ values, and the red solid circle shows the maximizer of the EI criterion.}\label{fig:ex2_1} 
\end{figure}

\textbf{Example 2}. Consider a computer model with two-dimensional input variables $\bx = (x_1,x_2)$, and the output given by
 \begin{equation} \label{eq:2d}
y(\bx) = [1 + (4x_1 + 4x_2 + 1)](3 + 192x_1x_2) , 0 \leq x_1 \leq 1, 0 \leq x_2 \leq 1.
\end{equation}
Suppose a maximin Latin hypercube design of 10 points is generated and the corresponding responses are collected from the computer model. First, we consider searching for the next follow-up design point for estimating only one contour at the level $a = 300$. Figure~\ref{fig:ex2_1}  shows these 10 design points, the inputs with $I(\bx)>0$, and the maximizer of the EI criterion for contour estimation in (\ref{eq:ei_original}) for the candidate set on a regular $100\times 100$ rectangular grid.  

\begin{figure}[h!] \centering
\includegraphics[width=0.6\textwidth]{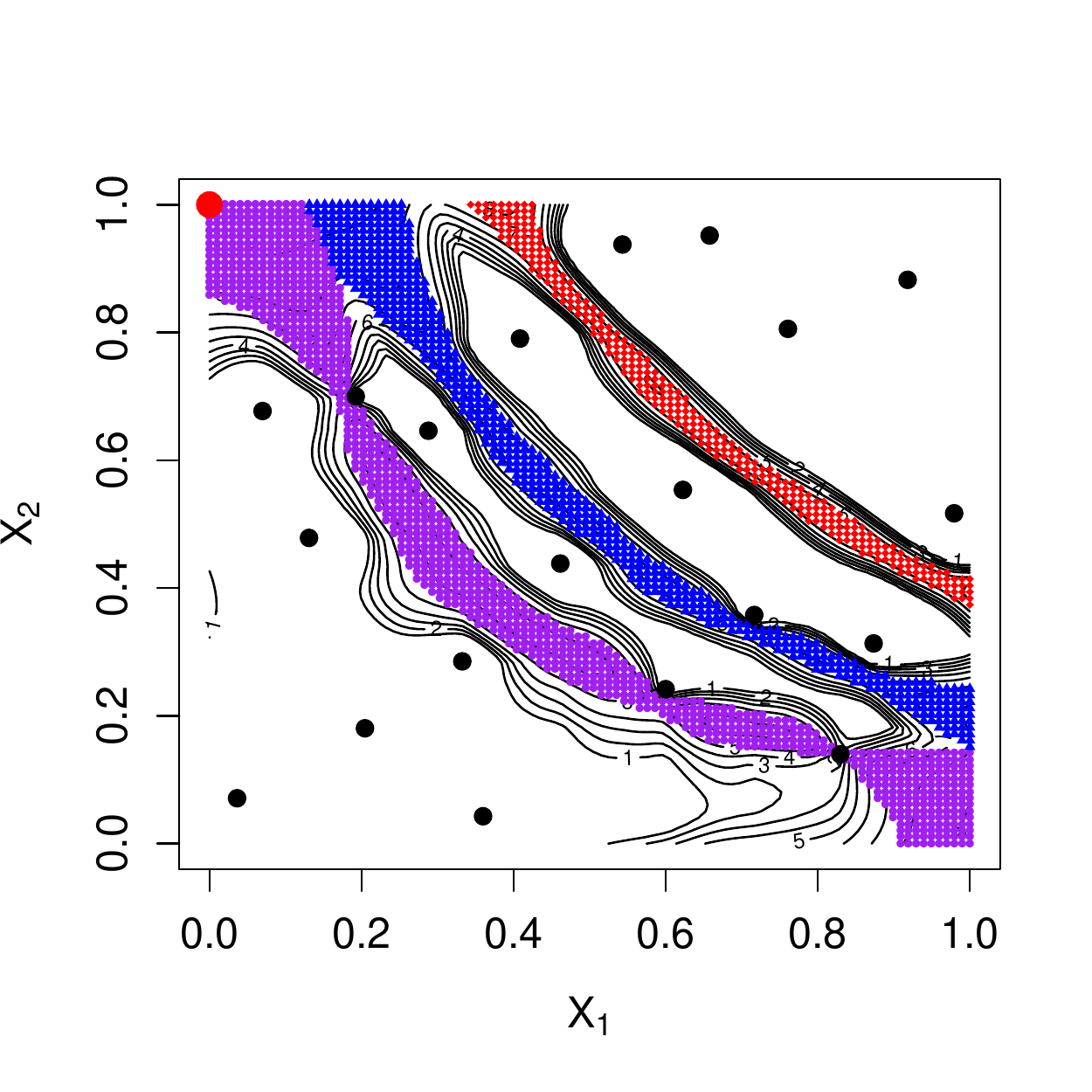}
\caption{Illustration of the follow-up point selection method using the MC criterion (at levels $a_1=150$, $a_2=300$,  $a_3=600$) for the computer model (\ref{eq:2d}). The black solid circles denote $10$ training points. The purple solid circles, blue triangles, red diamonds represent improvement around the three contour levels respectively. The contour lines display $\log(E[I(\bx)])$,  and the red solid circle represents the maximizer of the MC criterion in (\ref{eq:ei_mc}).}\label{fig:ex2_2}
\end{figure}

Next, we consider the simultaneous estimation of three contours at levels $a_1 = 150$, $a_2 = 300$ and $a_3 = 600$ using the MC criterion in (\ref{eq:ei_mc}).  Figure~\ref{fig:ex2_2}  shows the inputs from the same  $100\times 100$ grid candidate set that achieve non-zero improvement (\ref{eq:imp}),  and the point that 
maximizes the MC criterion. This point in red in Figure~\ref{fig:ex2_2}  is in fact from the set of the points that yield non-zero improvement around the contour level $a_1=150$.

\begin{figure}[h!]
\centering
\subfigure[Adding 1st follow-up point]{\label{fig:3a} \includegraphics[width=0.31\textwidth]{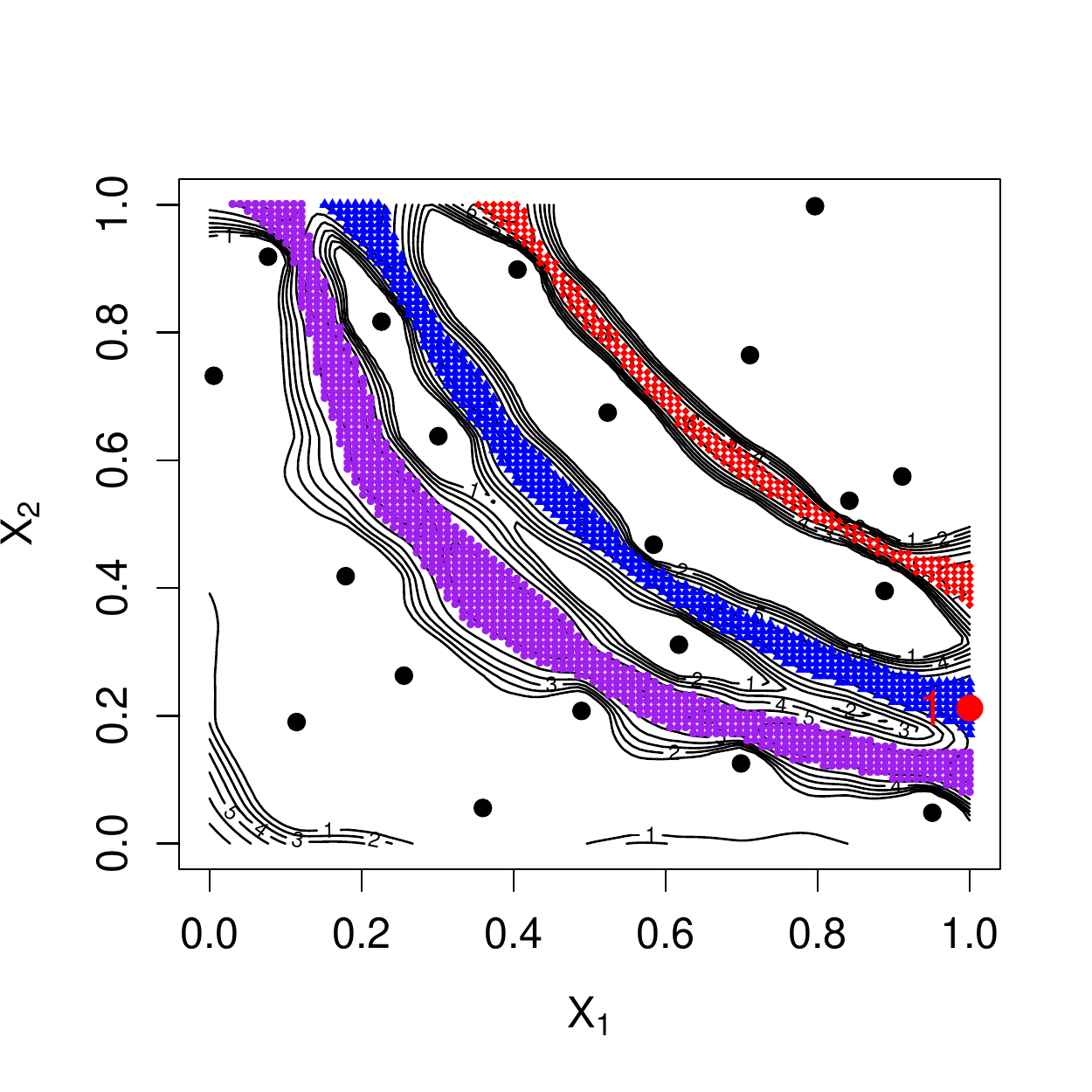}}
\subfigure[After adding 5 points]{\label{fig:3b}  \includegraphics[width=0.31\textwidth]{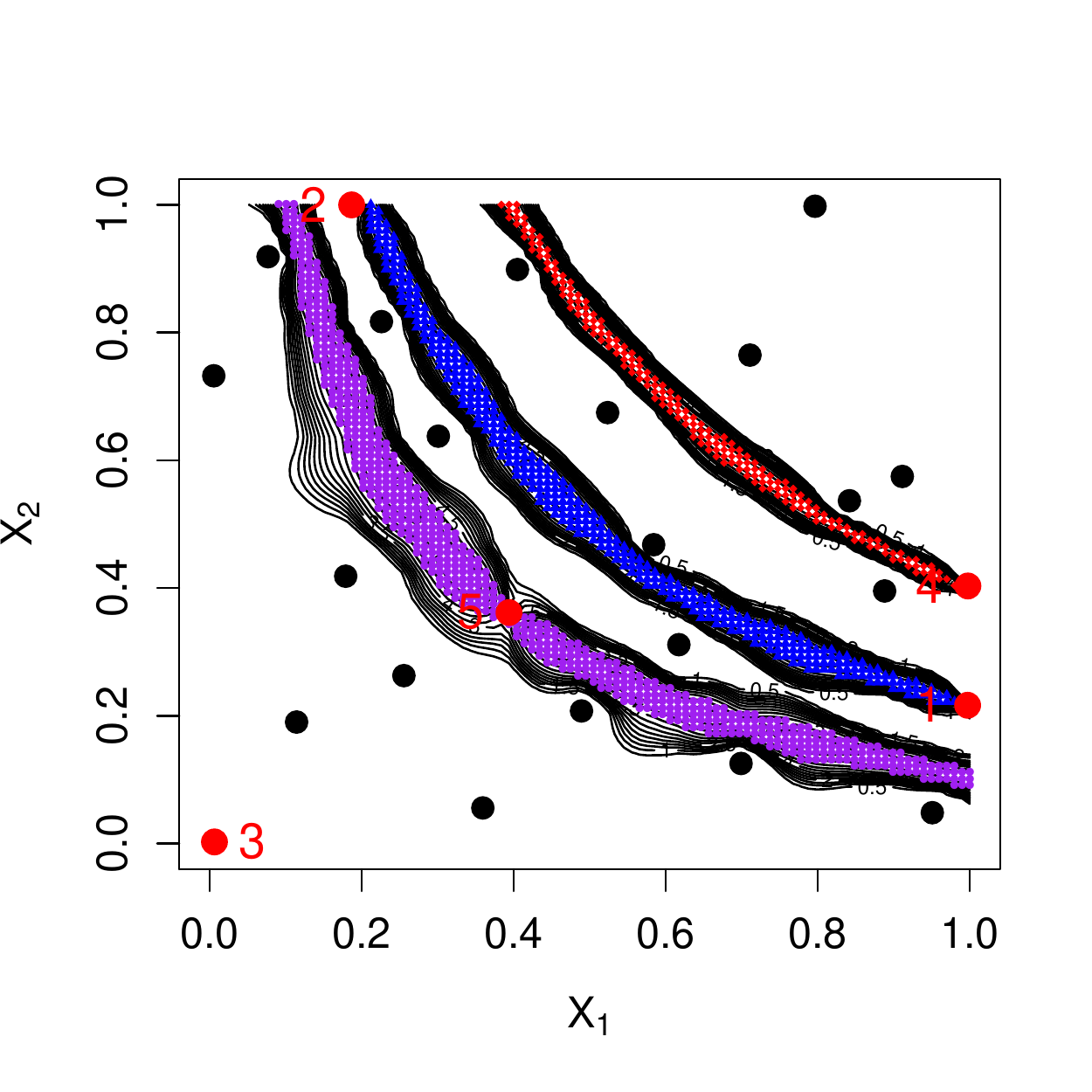}}
\subfigure[After adding 15 points]{\label{fig:3c}  \includegraphics[width=0.31\textwidth]{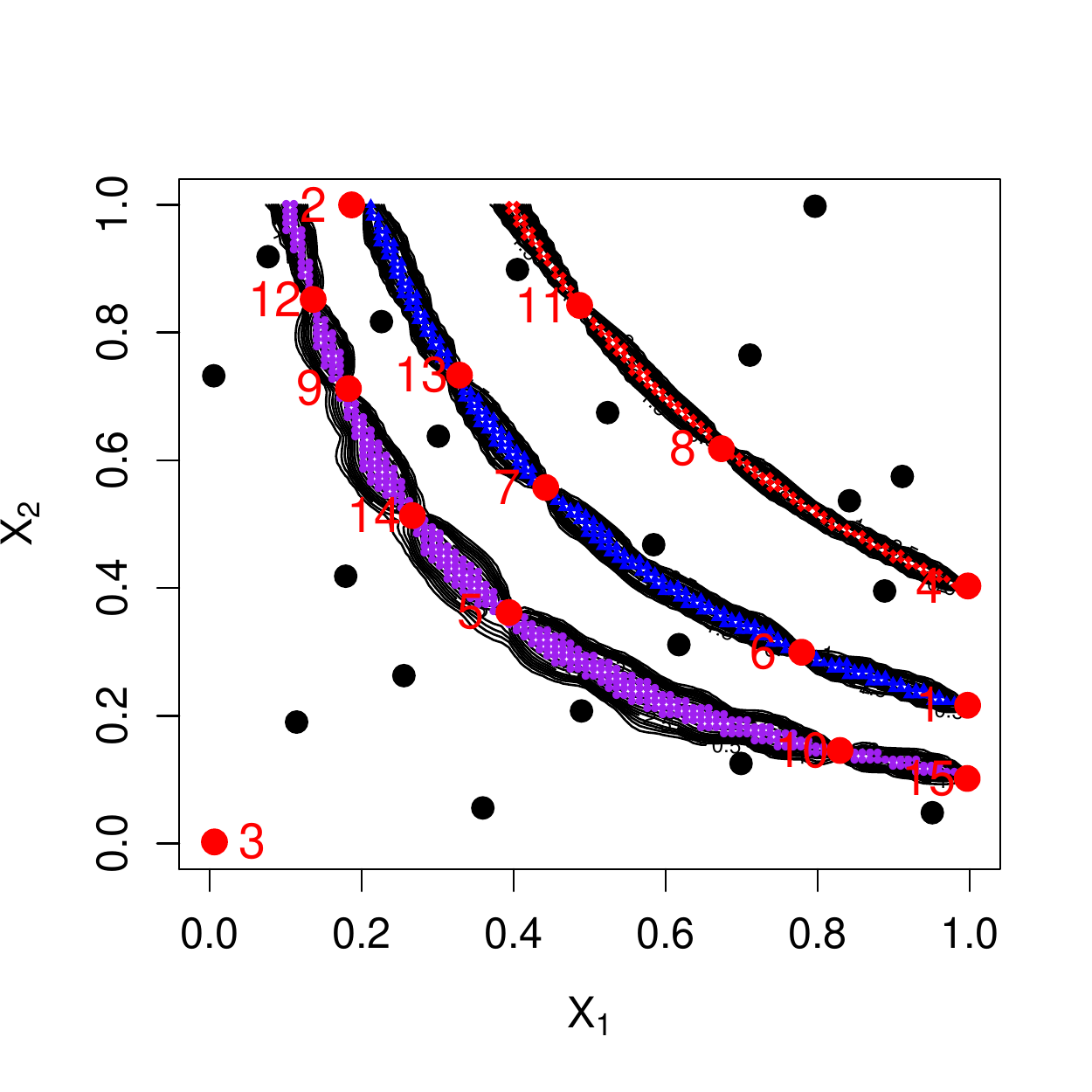}}
\subfigure[After adding 25 points]{\label{fig:3d}  \includegraphics[width=0.31\textwidth]{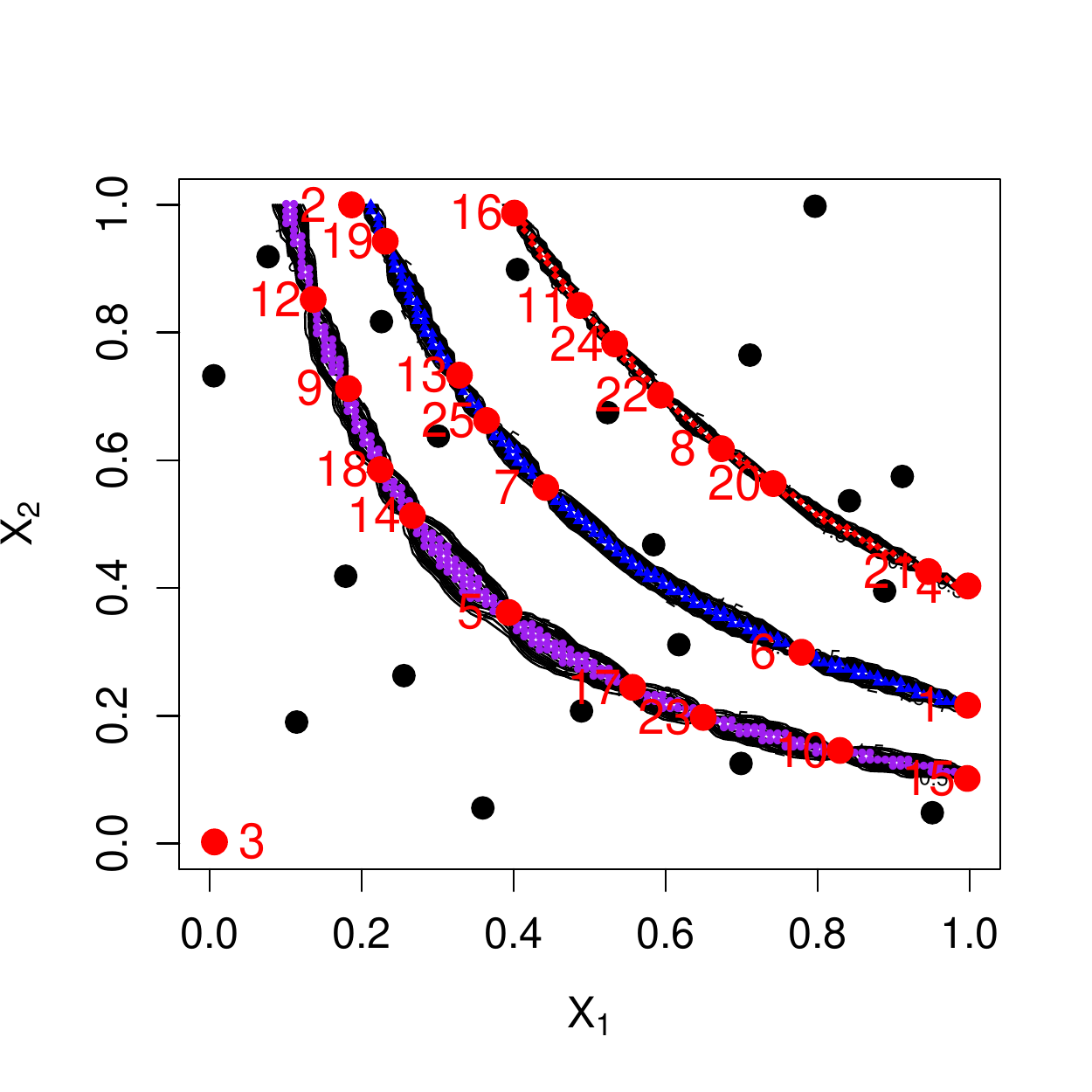}}
\subfigure[After adding 30 points]{\label{fig:3e}  \includegraphics[width=0.31\textwidth]{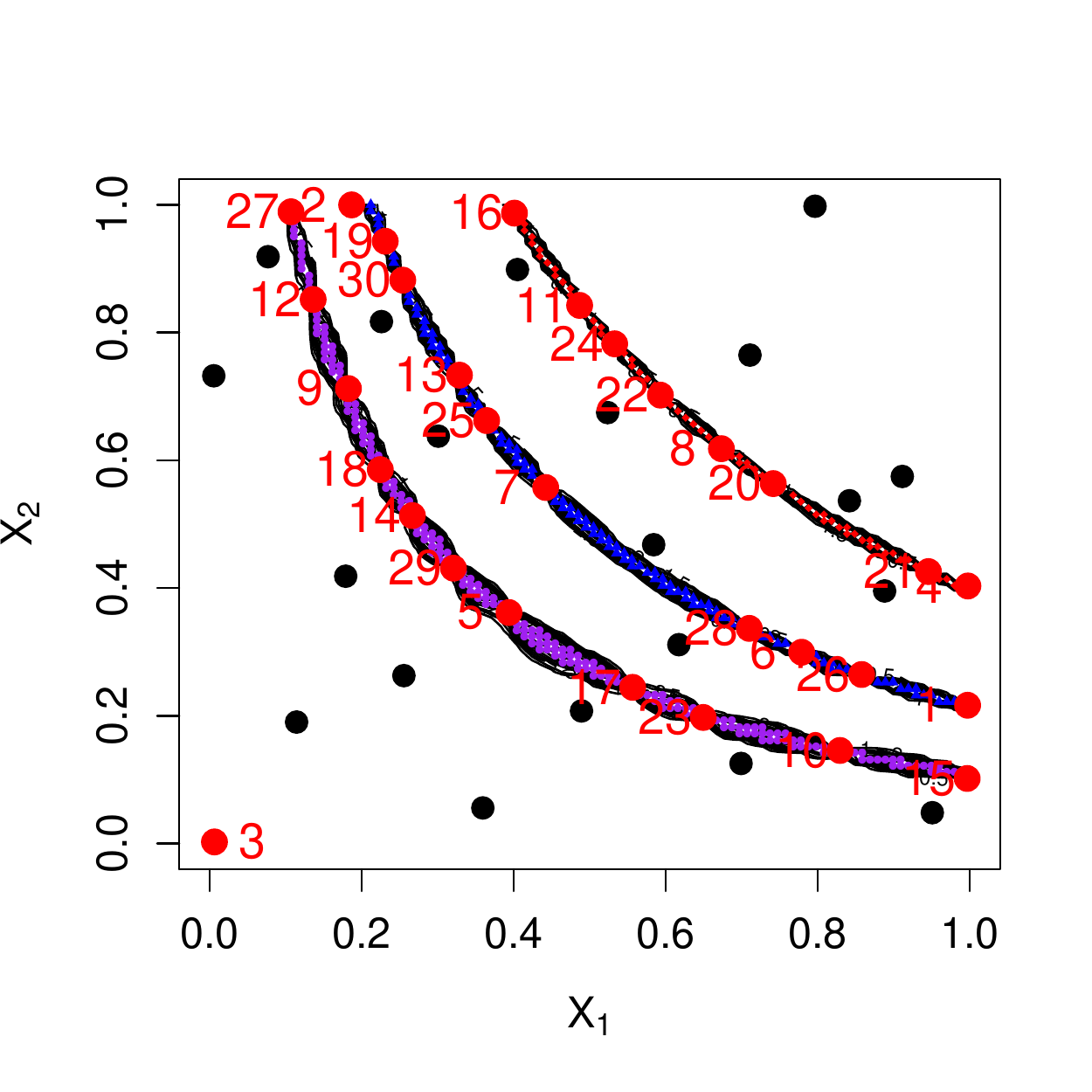}}
\subfigure[Running estimate of accuracy]{\label{fig:3f}  \includegraphics[width=0.31\textwidth]{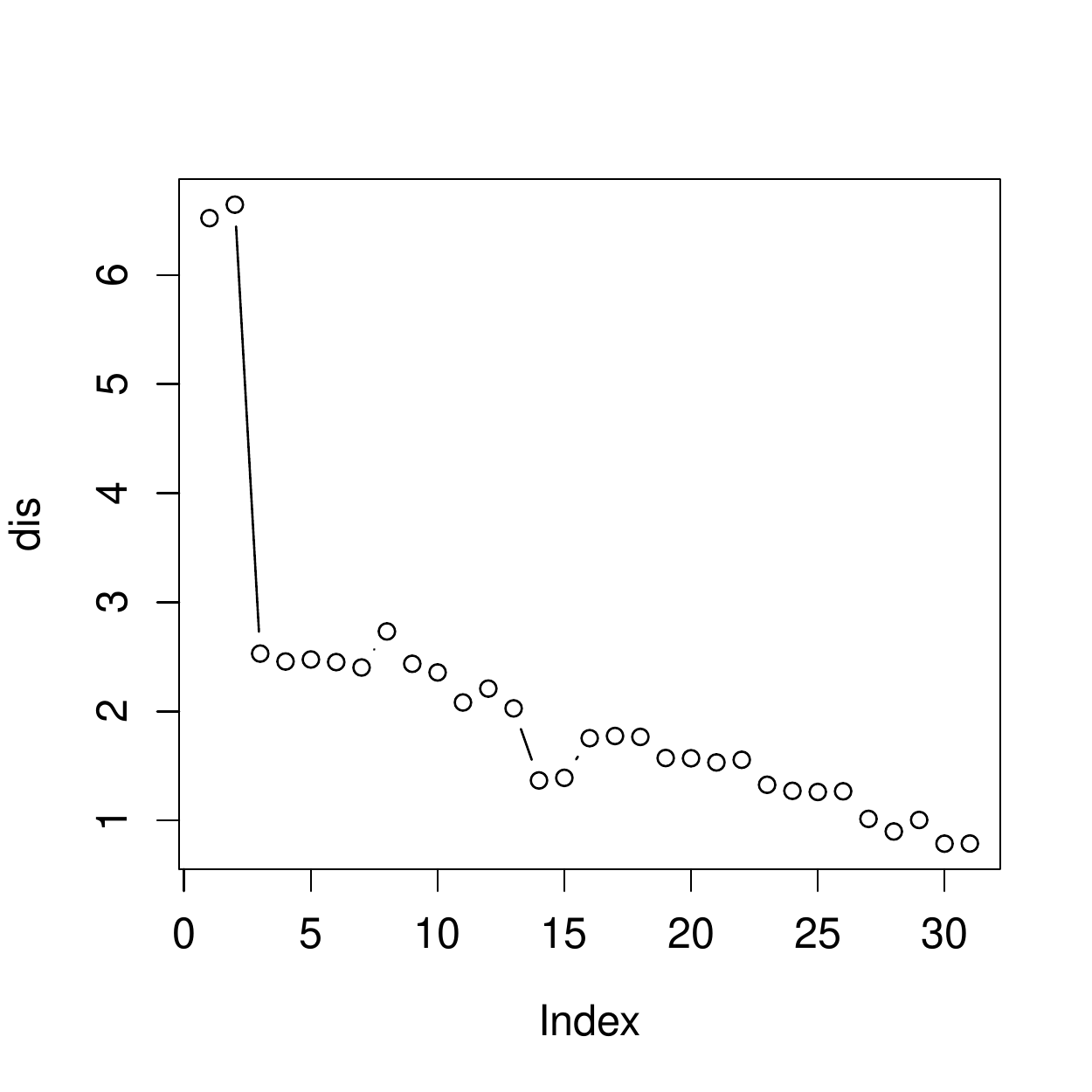}}
\caption{Illustration of the MC criterion for contour levels $a_1=150$, $a_2=300$ and $a_3=600$ with $n_0=10$ initial design points and $30$ follow-up points . The accuracy in Panel (f) is measured by the squared distance between the estimated contour after adding $i$-follow-up points and the true contour.}\label{fig:ex2_3}
\end{figure}

Figure~\ref{fig:ex2_3}  illustrates the complete sequential design scheme with $10$ initial design points and $30$ follow-up design points for simultaneously estimating three contours at levels $a_1=150$, $a_2=300$ and $a_3=600$.  The red points are the new follow-up points and the label corresponds to the order the point is added.  The last panel displays the squared distance between the estimated contour and the true contour at each stage. It can be observed from Figure~\ref{fig:ex2_3}  that the estimated uncertainty bands around the three contours become narrower and more accurate. It can also be seen that more points are added to estimate the contour level $a_1=150$ than to estimate the other two contours. Some points such as the second and third points are away from the contour bands.

It is clear from the two examples that the resulting designs do not have the conventional space-filling property. This is desirable as the objective is an overall good fit of the response surface, and not to explore the input space. However, as illustrated in Figure~\ref{fig:ex2_3}, a significant fraction of design points tend to line up on the pre-specified contours, which could lead to biased designs if $a_1, ..., a_k$ are not chosen appropriately. Next we propose an efficient method of selecting contour levels. 


 \section{Sequential Estimation of Contours for Global Fitting}

In this section we propose a new approach for choosing the follow-up design points.
Different from the previous section where the simultaneous estimation of multiple contours was used for global fitting, we adopt the EI criterion for estimating only one contour level at each stage. More importantly, we propose using different contour levels at different stages. That is, we sequentially find the design points that maximize the criterion (\ref{eq:ei_original}) with a different value of $a$ at each stage. The important issue is how to choose the contour level at each stage.  We propose the following way to choose such contour level in an automatic way.

Suppose at stage $j$, the training data are $\{ (\bx_i, y_i), i=1,\ldots,n)\}$ and the corresponding emulator gives the predictive distribution as $y(\bx) \sim N(\hat{y}(\bx), s^2(\bx))$ for any input $\bx$. Let  the candidate set for the next follow-up point be $\bx^*_1, \ldots, \bx^*_m$ and
$$ \bx^*_{opt} = \underset{1 \leq i \leq m}{\arg \max}\ s^2(\bx^*_i).$$
Then, we choose the contour level at stage $j$ as $a_j = \hat{y}(\bx^*_{opt})$.  In other words, at each stage, we set the contour level to be the fitted response that has maximum predictive variance. This is to encourage exploring the area with maximum uncertainty.

\begin{figure}[h!]
\centering
\ \includegraphics[width=0.65\textwidth]{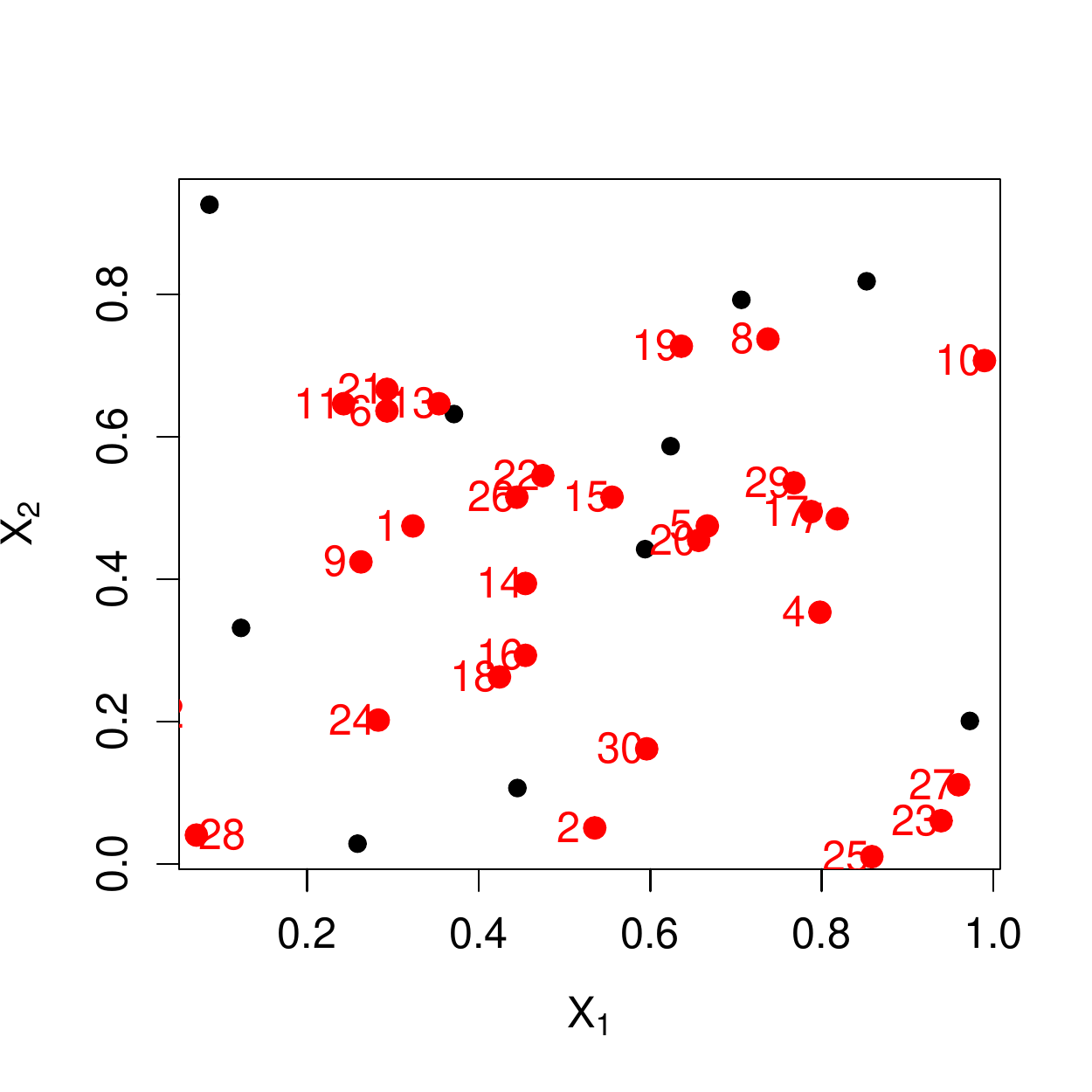}
\caption{Illustration of the sequential contour estimation-based EI criterion for global fitting   with $n_0=10$ and 30 added points.}\label{fig:ex2_4}
\end{figure}

\textbf{Example 2  (contd.)}  Consider finding a design for global fitting of the computer simulator in Example 2. The procedure starts with an initial design of size $n_0=10$ obtained via maximin Latin hypercube sampling and $n-n_0=30$ follow-up points are chosen as per the proposed sequential strategy. Figure~\ref{fig:ex2_4} displays the follow-up design points found by the proposed method, i.e., the sequential contour estimation-based EI criterion.

Note that the resulting design is more randomized as compared to a systematic layout of points on the contour lines shown in Figure~\ref{fig:ex2_3}. Again, the design is not completely space-filling and it has some pairs of close-by points.



\section{Simulated Examples}

In this section, we conduct a simulation study to  demonstrate the effectiveness of the proposed sequential design approaches. Specifically, we compare the proposed approaches with the following methods:
\begin{itemize}
\item[(a)] a one-shot maximin Latin hypercube design;
\item[(b)] the sequential $D$-optimal design in the \proglang{R} package \pkg{tgp};
\item[(c)] the sequential approach by Lam and Notz (2008); 
\item[(d)] the sequential minimum energy design in Joseph et al. (2015); 
\item[(e)] the proposed multiple contours estimation-based criterion in Section 3;
\item[(f)] the proposed sequential contour estimation-based criterion in Section 4.
\end{itemize} 
For approach (e), we use the 10 contour levels, that is, $k$ is set to be 10. 
 These methods are denoted by `maximinLHD', `tgp',`EIGF', `SMED', `MC$\_$10', and `SC$\_$var'.

Several criteria can be used to evaluate the performance of different design approaches in comparison. We adopt the root mean square prediction error (RMSPE) given by
\begin{equation}\label{eq:rmse}
\hbox{RMSPE} =  \sqrt{\frac{1}{|\mathcal{X}_{pred}|}\sum_{\bx \in \mathcal{X}_{pred}} (\hat{y}(\bx) - y(\bx))^2},
\end{equation}
where $\hat{y}(\bx)$ and $y(\bx)$ are the predicted response and the true response at the new input $\bx$ in the hold-out set $\mathcal{X}_{pred}$.
Another criterion we use is the maximum error provided by \begin{equation}\label{eq:nse}
\hbox{Maximum error} = \max_ {\bx \in \mathcal{X}_{pred}} | \hat{y}(\bx) - y(\bx)|.
\end{equation} 

For each example below, the initial design for sequential designs is a maximin Latin hypercube design of $n_0$ runs generated using the \proglang{R} package \pkg{SLHD} (Ba, 2015). The model fitting is implemented using the default setting of the function \code{GP\_fit} in the \proglang{R} package \pkg{GPfit}. The test data is a random Latin hypercube design of $1000d$ points where $d$ is the number of input variables. The parameter $\alpha$ in `MC$\_$10' and `SC$\_$var' is set to be 2.

\begin{figure}[h!]
\centering
\subfigure[]{\label{fig:ex3a} \includegraphics[width=0.7\textwidth]{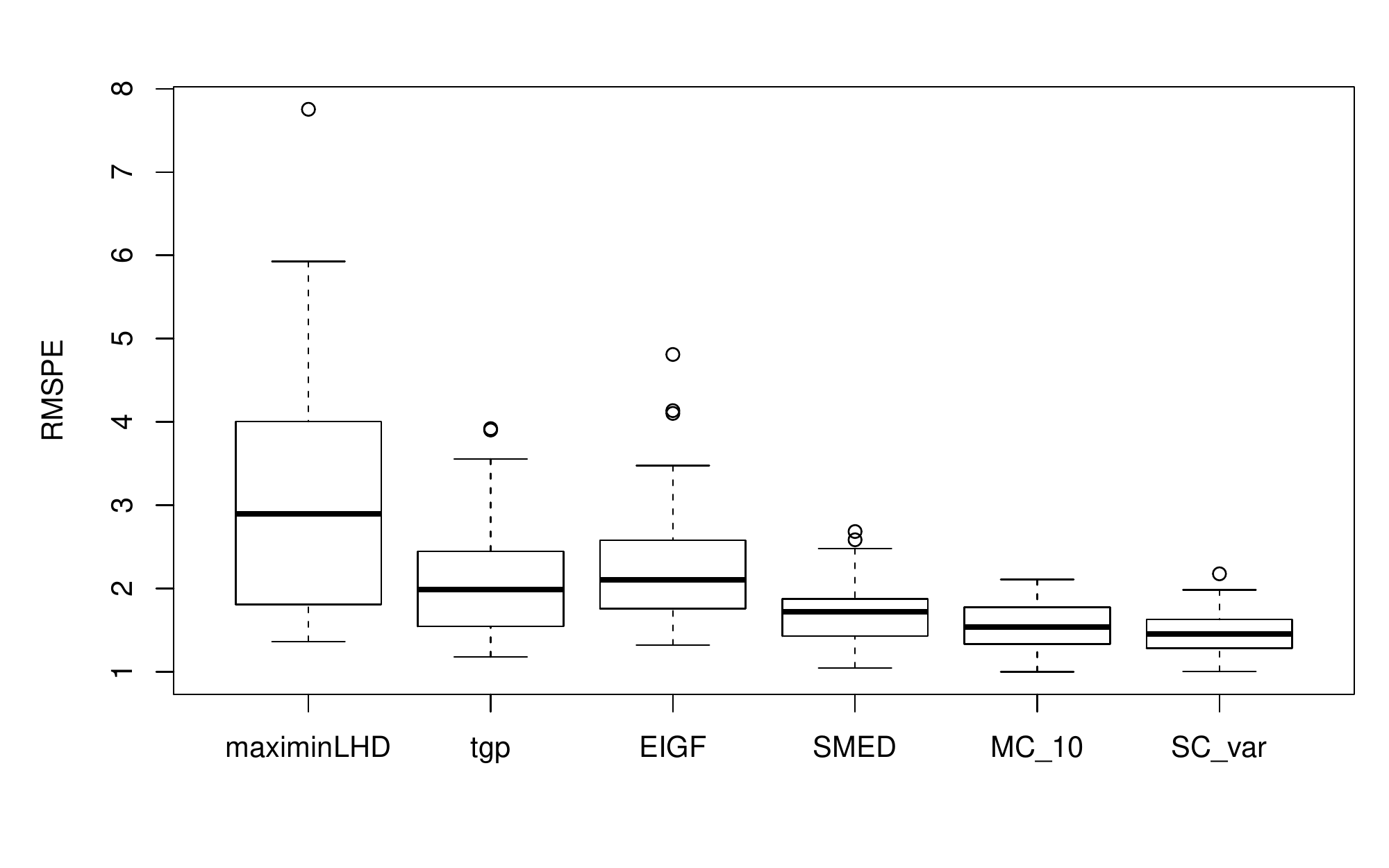}}
\subfigure[]{\label{fig:ex3b}  \includegraphics[width=0.7\textwidth]{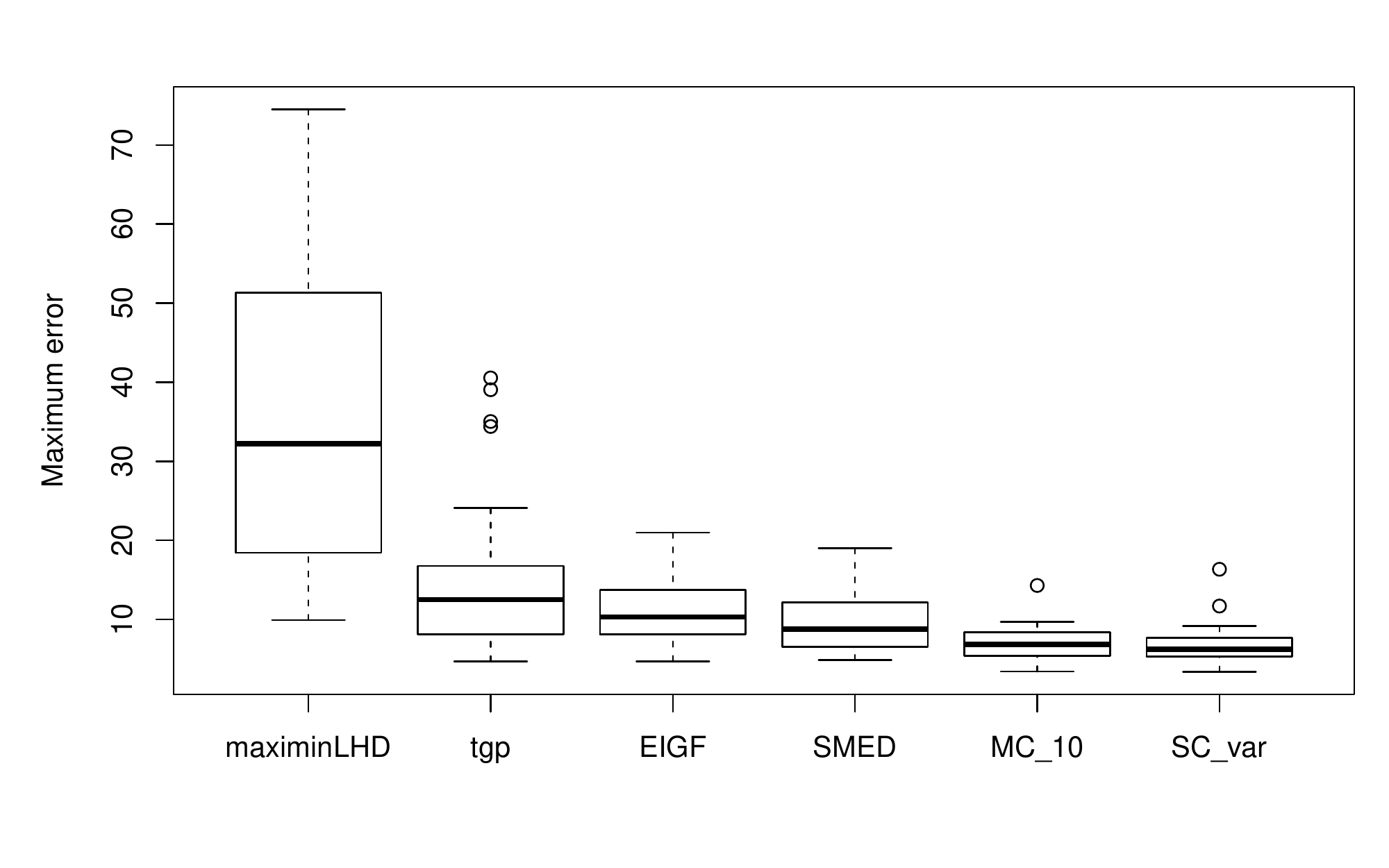}}
\caption{The boxplots of RMSPEs and maximum errors of the methods `maximinLHD', `tgp', `EIGF', `SMED', `MC$\_10$', and `SC$\_$var' for the computer model in (\ref{eq:ex3}) with $n_0=10$ and 20 added points over 50 simulations.}\label{fig:ex3}
\end{figure}

\noindent \textbf{Example 3}. We consider computer model with two input variables $x_1$ and $x_2$,
\begin{eqnarray}\label{eq:ex3}
y  &=&  \left(x_2 - \frac{5.1}{4\pi^2} x_1^2 + \frac{5}{\pi}x_1-6\right)^2 \\ \nonumber
     && +\ 10 \left(1- \frac{1}{8\pi}\right)\hbox{cos}(x_1) + 10,\  -5 \leq x_1 \leq 10, 0 \leq x_2 \leq 15.
\end{eqnarray}
This model is known as Branin function (Dixon and Szego, 1978). We use $n_0=10$ initial design points. The total run size budget is 30.  Figure~\ref{fig:ex3} displays the boxplots of RMSPEs and maximum errors of the different design approaches over 50 simulations. The results show that in this example the one-shot approach `maximinLHD' is the worst while the approaches `MC$\_$10' and `SC$\_$var'  are comparably better than the others.

\begin{figure}[h!]
\centering
\subfigure[]{\label{fig:ex4a} \includegraphics[width=0.7\textwidth]{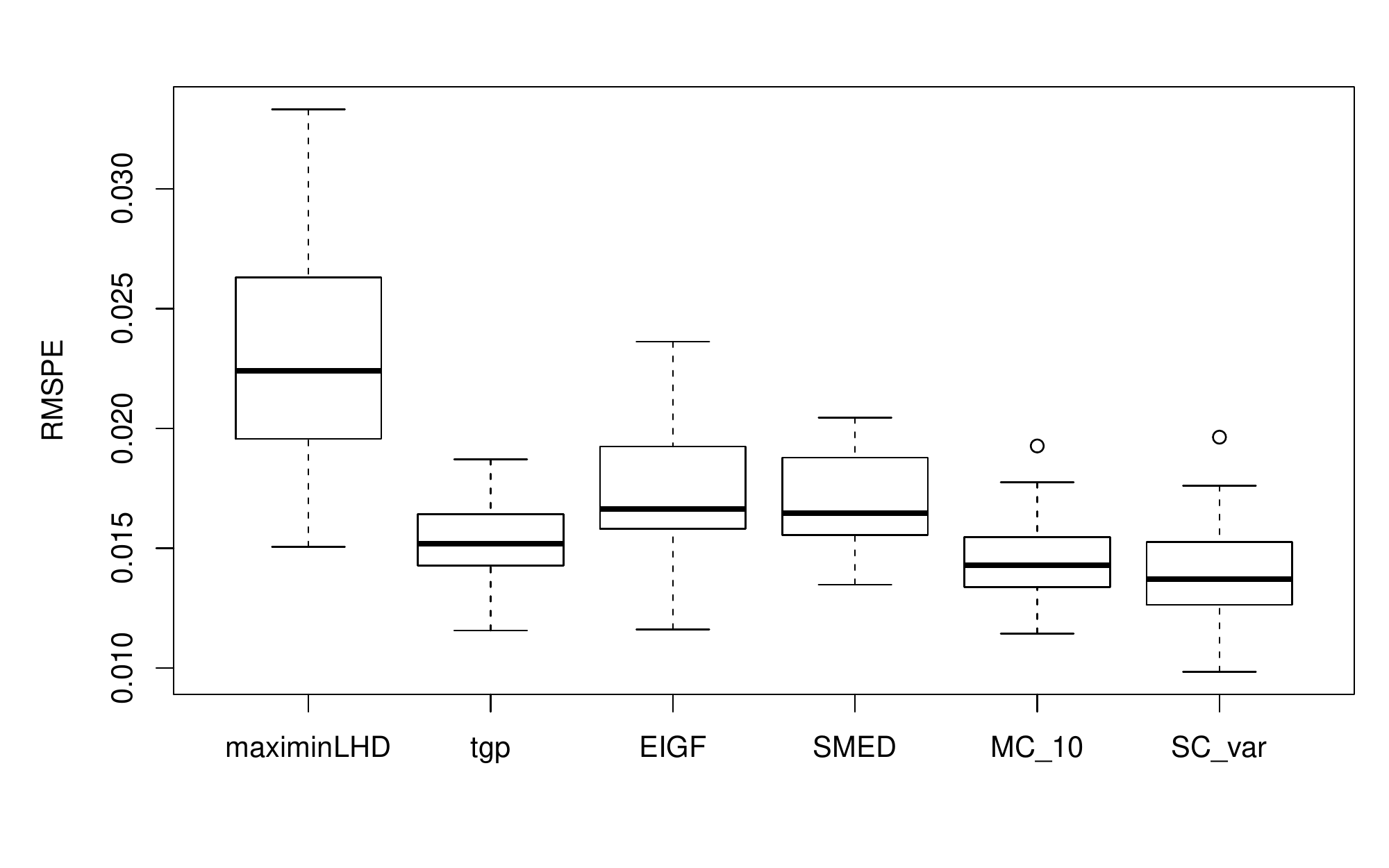}}
\subfigure[]{\label{fig:ex4b}  \includegraphics[width=0.7\textwidth]{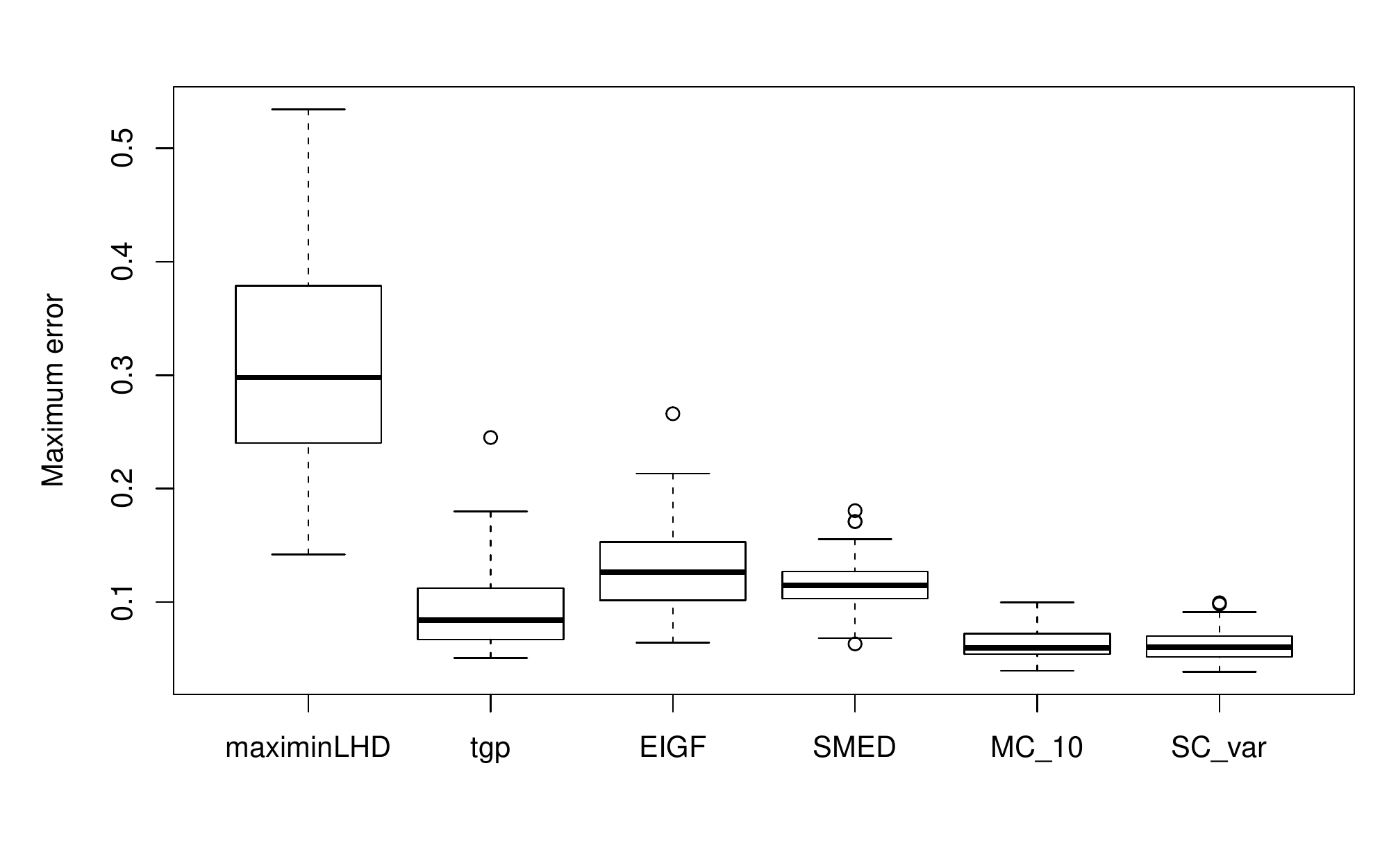}}
\caption{The boxplots of RMSPEs and maximum errors of the methods `maximinLHD', `tgp', `EIGF', `SMED', `MC$\_10$', and `SC$\_$var' for the computer model in (\ref{eq:ex4}) with $n_0=20$ and 40 added points over 50 simulations.}\label{fig:ex4}
\end{figure}

\noindent \textbf{Example 4}. We consider computer model with three input variables $x_1$, $x_2$ and $x_3$,
\begin{equation}\label{eq:ex4}
y =  \prod_{i=1}^3x_i,   0 \leq x_i \leq i,  \hbox{ for}, i=1,\ldots,3.
\end{equation}
We use $n_0=20$ initial design points. The total run size budget is 60.  Figure~\ref{fig:ex4} displays the boxplots of RMSPEs and maximum errors of the different design approaches over 50 simulations. Again, here the one-shot approach `maximinLHD' is the worst. The approaches `tgp', `MC$\_$10' and `SC$\_$var'  are comparably better than the others in terms of RMSPE and the proposed approaches `MC$\_$10' and `SC$\_$var'  are significantly better than the others in terms of maximum errors. 
  
\begin{figure}[h!]
\centering
\subfigure[]{\label{fig:ex5a} \includegraphics[width=0.7\textwidth]{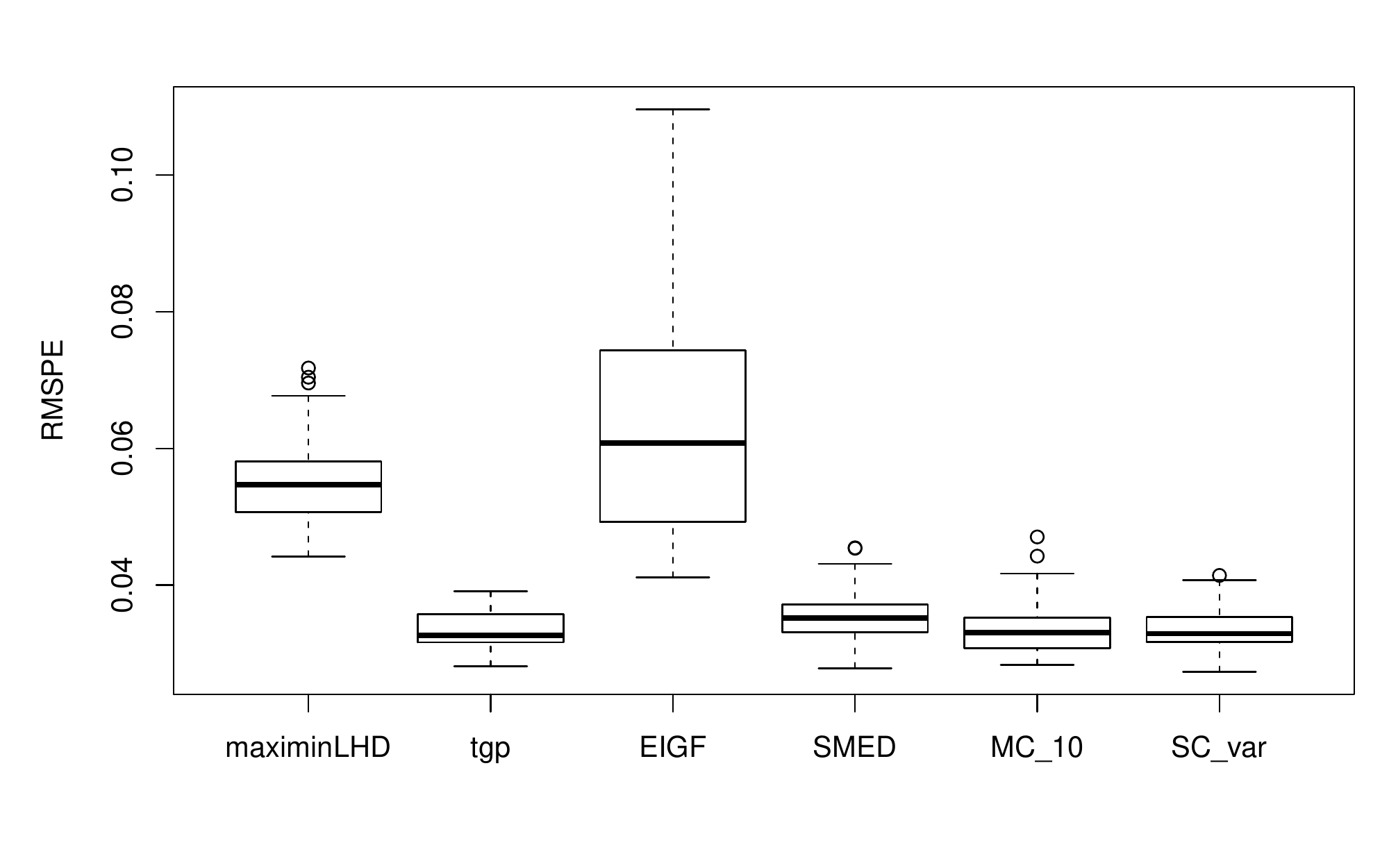}}
\subfigure[]{\label{fig:ex5b}  \includegraphics[width=0.7\textwidth]{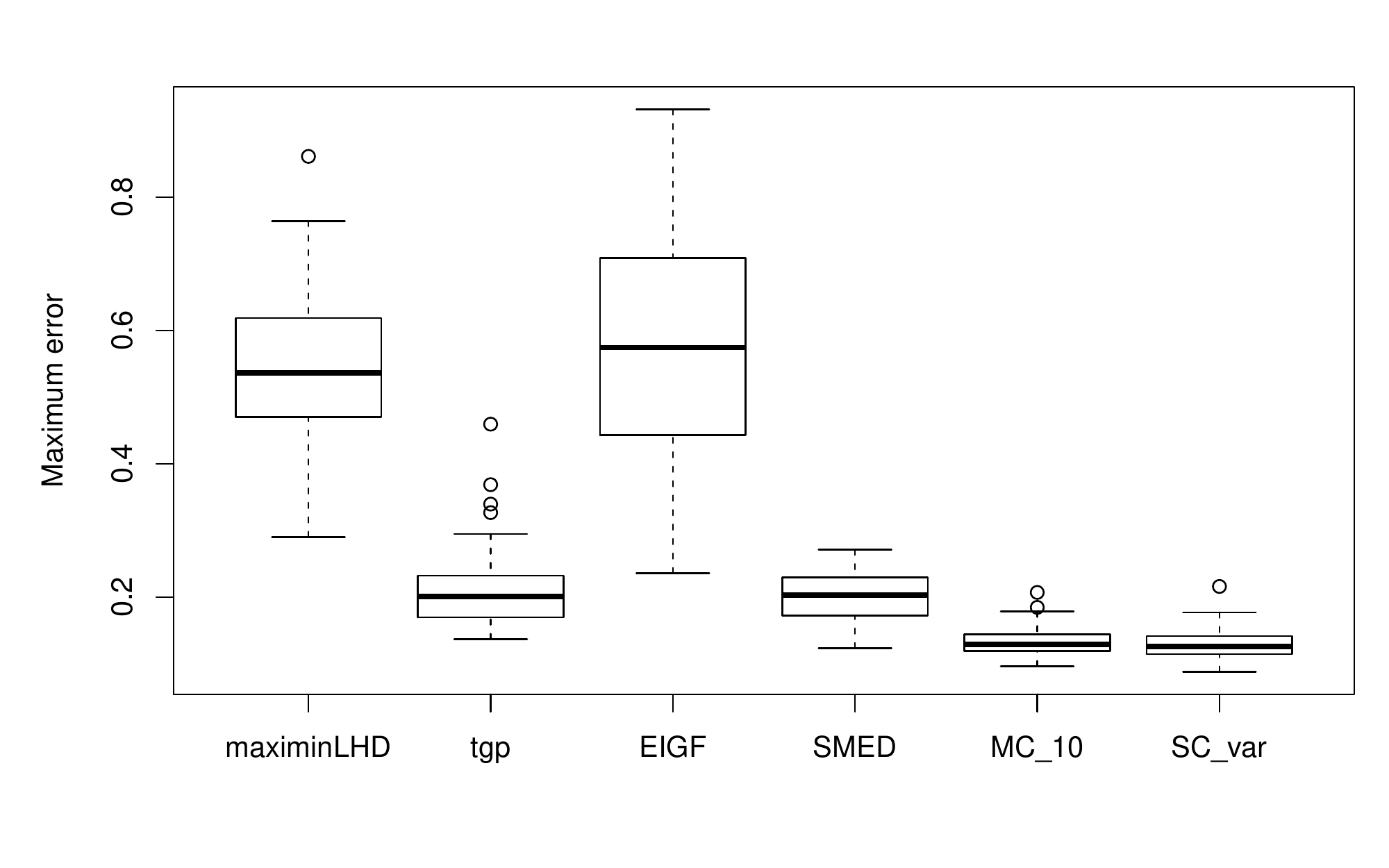}}
\caption{IThe boxplots of RMSPEs and maximum errors of the methods `maximinLHD', `tgp', `EIGF', `SMED', `MC$\_10$', and `SC$\_$var' for the computer model in (\ref{eq:ex5}) with  with $n_0=27$ and 53 added points over 50 simulations.}\label{fig:ex5}
\end{figure}

\noindent \textbf{Example 5}. We consider computer model with four input variables $x_1$, $x_2$, $x_3$ and $x_4$,
\begin{equation}\label{eq:ex5}
y =   x_1x_2 + x_3^2x_4^2,    -1 \leq x_i \leq 1, \hbox{ for}, i=1,\ldots,4. 
\end{equation}
We use $n_0=27$ initial design points. The total run size budget is 80.  Figure~\ref{fig:ex5} displays the boxplots of RMSPEs and maximum errors of the different design approaches over 50 simulations. Here, the approach `EIGF' is the worst followed by the approach `maximinLHD' in terms of both criteria. The performance of the other four approaches are similar based on RMSPE. However, based on the maximum error, the proposed approaches give more accurate predictions. 
  
\section{Concluding Remark}

In this article we have developed two sequential design approaches
for  accurately predicting a complex computer code. The approaches are based on the expected improvement criteria for simultaneously or sequentially estimating contours. 
We used a Gaussian process (GP) model as a surrogate for the computer simulator, which is an integral component of the proposed criteria for identifying the follow-up trials. Numerical examples are given to demonstrate that the proposed approaches can significantly outperform the existing approaches.

Note that if some other surrogate is used instead of GP model, then also the key ideas like formulation of improvement function and sequential estimation of contour levels can be retained. Of course, the resultant expected improvement criteria would change, and in fact, one may not even end up with a closed form expression of the final design criterion for selecting follow-up points. Future work also include the application of the proposed contour estimation-based sequential design approaches for global fitting for computer experiments with both qualitative and quantitative factors (Deng et al., 2017) and dynamic computer experiments (Zhang, Lin and Ranjan, 2018).

\section*{References}
\begin{description}
\item[] Ba, S.\  (2015). SLHD: Maximin-Distance (Sliced) Latin Hypercube Designs. R package version 2.1-1.

\item[] Bayarri, M. J., Berger, J. O., Calder, E. S., Dalbey, K., Lunagomez, S., Patra, A. K., Pitman, E. B., Spiller, E. T., and Wolpert, R. L. (2009).\  ``Using statistical and computer models to quantify volcanic hazards,'' \emph{Technometrics}, 51, 402--413.

\item[] Bower,  R.  G.,  Benson,  A.  J.,  et  al.\  (2006).   ``The  broken  hierarchy  of  galaxy  formation,'' \emph{Monthly Notices of the Royal Astronomical Society}, 370, 645--655.

\item[] Chipman, H., Ranjan, P., and Wang, W.\ (2012). ``Sequential design for computer experiments with a flexible Bayesian additive model,''{\em Canadian Journal of Statistics}, 40(4), 663-678.

\item[] Dancik, G.\ (2018). mlegp: Maximum Likelihood Estimates of Gaussian Processes. R package version
3.1.7.

\item[] Deng, X., Lin, C.D., Liu, K.W., and Rowe, R.K.\ (2017).  ``Additive Gaussian process for computer models with qualitative and quantitative factors,'' {\em Technometrics}, 59, 283-292.

\item[] Dixon, L. C. W. and Szego, G. P.\ (1978). ``The global optimization problem: an introduction, ''{\em Towards Global Optimization}, 2, 1-15.

\item[] Fang, K. T., Li, R. and Sudjianto, A.\ (2005). {\it Design and Modeling for Computer Experiments},  New York: Chapman$\&$Hall/CRC Press.
 
\item[] Gramacy, R.B. and Lee, H.K.H.\ (2008). ``Bayesian treed Gaussian process models with an application to computer modeling,''{\em Journal of the American Statistical Association}, 103(483), 1119--1130.

\item[] Gramacy, R.B. and Lee, H.K.H.\ (2009), ``Adaptive design and analysis of supercomputer experiments,'' \emph{Technometrics}, 51(2), 130--145. 

 \item[] Gramacy, R. B. and  Lee, H.K.\ (2012). ``Cases for the nugget in modeling computer experiments.,''{\em Statistics and Computing}, 22(3), 713--722.
 
\item[] Gramacy, R. B. and Taddy, M.A.\ (2016). tgp: Bayesian Treed Gaussian Process Models. R package version
2-4-14.

\item[] Greenberg, D.\ (1979). ``A numerical model investigation of tidal phenomena
in the Bay of Fundy and Gulf of Maine,'' \emph{Marine Geodesy}, 2, 161--187.

 \item[] Gu, M. Palomo, J., and Berger, J.\ (2018).  RobustGaSP: Robust Gaussian Stochastic Process Emulation. R package version 0.5.6.

\item[] Han, G., Santner, T. J., Notz, W. I.,  and Bartel, D. L.\ (2009).  ``Prediction for computer experiments having quantitative  and qualitative input variables,'' {\em Technometrics},  51, 278--288.

\item[] Iman, R. L. and Conover, W. J.\ (1982). ``A distribution-free approach to inducing rank correlation among input variables,'' \emph{Communication in Statistics Part B-Simulation and Computing}, 11, 311--334.

\item[] Jesus, P., Garcia-Donato, G., Paulo, R., Berger, J., Bayarri, M., and Sacks, J. \ ( 2019). SAVE: Bayesian Emulation, Calibration and Validation of Computer Models. R package version 1.0.

\item[] Johnson, M., Moore, L. and Ylvisaker, D.\ (1990). ``Minimax and maximin distance design,'' \emph{Journal of Statistical Planning and Inference,} 26, 131--148.

\item[] Jones, D. R., Schonlau, M., and Welch, W. J.\ (1998). ``Efficient global optimization of expensive black-box functions, ''{\em Journal of Global Optimization}, 13(4), 455--492.

\item[] Joseph, V.R., Dasgupta, T., Tuo, R., and Wu, C.J.\ (2015). ``Sequential exploration of complex surfaces using minimum energy designs,''{\em Technometrics}, 57, 64--74.

\item[] Joseph, V.R., Gul, E. and Ba, S., (2015). ``Maximum projection designs for computer experiments,'' {\em Biometrika}, 102, 371--380.

\item[] Joseph, V.R., and Hung, Y.\ (2008). ``Orthogonal-maximin Latin hypercube designs,''\emph{Statistica Sinica}, 18, 171--186.

\item[] Lam, C.Q. and Notz, W.I.\ (2008). ``Sequential adaptive designs in computer experiments for response surface model fit,'' {\em Statistics and Applications}, 6, 207--233.

\item[] Loeppky, J.L., Sacks, J., and Welch, W.J.\ (2009). ``Choosing the sample size of a computer experiment: a practical guide,''{\em Technometrics}, 51(4), 366--376.

\item[] MacDonald B, Ranjan P,  and Chipman H.\ (2015). GPfit: an R Package for Gaussian Process Model Fitting Using a New Optimization Algorithm. R package version 1.0-1.

\item[] McKay, M. D., Beckman, R. J.,  and Conover, W. J.\ (1979). ``A comparison of three methods for selecting values of input variables in the analysis of output from a computer code,'' {\em Technometrics},  21, 239--45.

\item[] Morris, M. D. and Mitchell, T. J. (1995). ``Exploratory designs for computer experiments, '' \emph{Journal of Statistical Planning and Inference,} 43, 381--402.

\item[] Owen, A.B., (1992). ``Orthogonal arrays for computer experiments, integration and visualization,'' {\em Statistica Sinica},  2, 439--452.

\item[] Palomo, J., Paulo, R., and Garcia-Donato, G.,\ (2015). ``SAVE: An r
  package for the statistical analysis of computer models''. \emph{Journal of
  Statistical Software}, 64(13), 1--23.

\item[] Roustant, O., Ginsbourger, D., and Deville, Y.\  (2018). DiceKriging: Kriging Methods for Computer Experiments. R package version 1.5.6.

\item[] Ranjan, P., Bingham, D., and Michailidis, G.\ (2008). ``Sequential experiment design for contour estimation from complex computer codes,''{\em Technometrics}, 50(4), 527--541. Errata (2011), \emph{Technometrics}, 53:1, 109-110.

\item[] Rasmussen, C.E. and  Williams., C.K.I. (2006). \emph{Gaussian Processes for Machine Learning}. The MIT Press, Cambridge, MA.

\item[] Santner, T.J., Williams, B. J.,  and Notz, W. I.\ (2003). {\it  The Design and Analysis of Computer Experiments}. New York: Springer.

\item[] Sacks, J., Schiller, S. B., and Welch W. J. (1989). ``Designs for computer experiments,'' {\em Technometrics}, 31, 41--47.

\item[] Sacks, J. and Schiller, S.\ (1988). Spatial designs. \emph{In Statistical Decision Theory and Related Topics IV} Vol. 2 (Gupta and Berger (eds.)), 385--399, Springer- Verlag, New York.

\item[] Sacks, J., Welch, W. J., Mitchell, T.J.,  and Wynn, H.P. (1989). ``Design and analysis of computer experiments,'' {\em Statistical Science}, 409--423.

\item[] Shewry, M.C. and Wynn, H.P.\ (1987). ``Maximum entropy sampling,'' {\em Journal of Applied Statistics}, 14(2), 165-170.

\item[] Tang, B. (1993). Orthogonal array-based Latin hypercubes. \emph{Journal of American Statistical Association}, 88, 1392--1397.

\item[] Zhang, R., Lin, C.D., and Ranjan, P. (2018). ``Local approximate Gaussian process model for large-scale dynamic computer experiments,'' {\em Journal of Computational and Graphical Statistics}, 27, 798--807.

 \end{description}

\end{document}